# Astrometry during the past 100 years

Erik Høg  Niels Bohr Institute, Copenhagen

2011.05.03:

ABSTRACT: The reports from 2008: "Astrometry and optics during the past 2000 years", are available at arXiv and at my website: www.astro.ku.dk/~erik/History.pdf . Here are now further contributions to the history of astrometry related to space astrometry. The development of photoelectric astrometry is followed from an experiment in 1925 up to the Hipparcos satellite mission in the years 1989-93. This period continues with my proposal in 1992 for CCD astrometry with a scanning satellite called Roemer, which led to the Gaia mission due for launch in 2013. Lectures on astrometry are described. - Further installments are planned.

2011.05.03:   Collection of reports from 2011. The following contains overview with summary and link to the reports Nos. 1-9 from 2008 and Nos. 10-13 from 2011. The reports are collected in two big file, see details on p.8.

## CONTENTS of Nos. 1-9 from 2008

# CONTENTS of Nos. 10-13 from 2011

# Overview with links to Nos. 1-9

No. 1 -  2008.05.27:

## Bengt Strömgren and modern astrometry:
## Development of photoelectric astrometry
## including the Hipparcos mission

ABSTRACT: Bengt Strömgren is known as the famous astrophysicist and as a leading figure in many astronomical enterprises. Less well-known, perhaps, is his role in modern astrometry although this is equally significant. There is an unbroken chain of actions from his ideas and experiments with photoelectric astrometry since 1925 over the new meridian circle in Denmark in the 1950s up to the Hipparcos and Tycho Catalogues published in 1997.
www.astro.ku.dk/~erik/Stroemgren.pdf
Contribution to IAU Symposium No. 254 in Copenhagen, June 2008: The Galaxy Disk in Cosmological Context – Dedicated to Professor Bengt Strömgren (1908-1987).

No. 1A - 2008.06.10:

## Bengt Strömgren and modern astrometry ... (Short version)
www.astro.ku.dk/~erik/StroemgrenShort.pdf
The same title as No. 1, but containing the short version posted at the symposium.

No. 2 - 2008.03.31:

## Lennart Lindegren's first years with Hipparcos
ABSTRACT: Lennart Lindegren has played a crucial role in the Hipparcos project ever since he entered the scene of space astrometry in September 1976. This is an account of what I saw during Lennart's first years in astrometry after I met him in 1976  when he was a young student in Lund.
www.astro.ku.dk/~erik/Lindegren.pdf

No. 3 – 2008.05.28:

## Miraculous approval of Hipparcos in 1980

ABSTRACT: The approval of the Hipparcos mission in 1980 was far from being smooth since very serious hurdles were encountered in the ESA committees. This process is illuminated here by means of documents from the time and by recent correspondence. The evidence leads to conclude that in case the approval would have failed, Hipparcos or a similar scanning astrometry mission would never have been realized, neither in Europe nor anywhere else.
www.astro.ku.dk/~erik/HipApproval.pdf

No. 4 -  2007.12.10:

## From the Roemer mission to Gaia

ABSTRACT: At the astrometry symposium in Shanghai 1992 the present author made the first proposal for a specific mission concept post-Hipparcos, the first scanning astrometry mission with CCDs in time-delayed integration mode (TDI). Direct imaging on CCDs in long-focus telescopes was described as later adopted for the Gaia mission. The mission called Roemer was designed to provide accurate astrometry and multi-colour photometry of 400 million stars brighter than 18 mag in a five-year mission. The early years of this mission concept are reviewed.
www.astro.ku.dk/~erik/ShanghaiPoster.pdf
Presented as poster at IAU Symposium No. 248 in Shanghai, October 2007. Only the first three pages appear in the Proceedings.

No. 5 -  2008.05.23, updated 2008.11.25.
Note in 2011: See further update in **www.astro.ku.dk/~erik/History2**

# Four lectures on the general history of astrometry

Overview, handout, abstracts at:   www.astro.ku.dk/~erik/Lectures.pdf
**Brief overview :**
  Lecture No. 1:

### Astrometry and photometry from space: Hipparcos, Tycho, Gaia

  The introduction covers 2000 years of astronomy from Ptolemy to modern times. The Hipparcos mission of the European Space Agency was launched in 1989, including the Tycho experiment. The Hipparcos mission and the even more powerful Gaia mission to be launched in 2011 are described.

  Lecture No. 2:

### From punched cards to satellites: Hipparcos, Tycho, Gaia

  A personal review of 54 years development of astrometry in which I participated.

  Lecture No. 3:

### The Depth of Heavens - Belief and Knowledge during 2500 Years

  The lecture outlines the understanding of the structure of the universe and the development of science during 5000 years, focusing on the concept of distances in the universe and its dramatic change in the developing cultural environment from Babylon and ancient Greece to modern Europe.

  Lecture No. 4, included on 2008.11.25:

### 400 Years of Astrometry: From Tycho Brahe to Hipparcos

  Four centuries of techniques and results are reviewed, from the pre-telescopic era up to the use of photoelectric astrometry and space technology in the first astrometric satellite, Hipparcos, launched by ESA in 1989. The lecture was presented as invited contribution to the symposium at ESTEC in September 2008: **400 Years of Astronomical Telescopes: A Review of History, Science and Technology.** The report submitted to the proceedings is included as No. 8 among "Contributions to the history of astrometry".

No. 6 – 2008.11.25:

## Selected astrometric catalogues

ABSTRACT: A selection of astrometric catalogues are presented in three tables for respectively positions, proper motions and trigonometric parallaxes. The tables contain characteristics of each catalogue to show especially the evolution over the last 400 years in optical astrometry. The number of stars and the accuracy are summarized by the weight of a catalogue, proportional with the number of stars and the statistical weight.
www.astro.ku.dk/~erik/AstrometricCats.pdf

No. 7 – 2008.11.25:

## Astrometric accuracy during the past 2000 years

ABSTRACT: The development of astrometric accuracy since the observations by Hipparchus, about 150 B.C., has been tremendous and the evolution has often been displayed in a diagram of accuracy versus time. Some of these diagrams are shown and the quite significant differences are discussed. A new diagram is recommended and documented.
www.astro.ku.dk/~erik/Accuracy.pdf
The two diagrams, Fig. 1a and 1b, in black/white and colour :
www.astro.ku.dk/~erik/AccurBasic.pdf      www.astro.ku.dk/~erik/AccuracyColour.jpg
www.astro.ku.dk/~erik/AccuracyBW.wmf      www.astro.ku.dk/~erik/AccuracyColour.wmf

No. 8 - 2008.11.25:

## 400 Years of Astrometry: From Tycho Brahe to Hipparcos

ABSTRACT: Galileo Galilei's use of the newly invented telescope for astronomical observation resulted immediately in epochal discoveries about the physical nature of celestial bodies, but the advantage for astrometry came much later. The quadrant and sextant were pre-telescopic instruments for measurement of large angles between stars, improved by Tycho Brahe in the years 1570-1590. Fitted with telescopic sights after 1660, such instruments were quite successful, especially in the hands of John Flamsteed. The meridian circle was a new type of astrometric instrument, already invented and used by Ole Rømer in about 1705, but it took a hundred years before it could fully take over. The centuries-long evolution of techniques is reviewed, including the use of photoelectric astrometry and space technology in the first astrometry satellite, Hipparcos, launched by ESA in 1989. Hipparcos made accurate measurement of large angles a million times more efficiently than could be done in about 1950 from the ground, and it will soon be followed by Gaia which is expected to be another one million times more efficient for optical astrometry.
www.astro.ku.dk/~erik/Astrometry400.pdf
Invited contribution to the symposium in Leiden in October 2008:
**400 Years of Astronomical Telescopes: A Review of History, Science and Technology**

No. 9 - 2008.11.25:

## 650 Years of Optics: From Alhazen to Fermat and Rømer

ABSTRACT: Under house arrest in Cairo from 1010 to 1021, Alhazen wrote his Book of Optics in seven volumes. (The caliph al-Hakim had condemned him for madness.) Some parts of the book came to Europe about 1200, were translated into Latin, and had great impact on the development of European science in the following centuries. Alhazen's book was considered the most important book on optics until Johannes Kepler's "Astronomiae Pars Optica" in 1604. Alhazen's idea about a finite speed of light led to "Fermat's principle" in 1657, the foundation of geometrical optics.
www.astro.ku.dk/~erik/HoegAlhazen.pdf
Contribution to the symposium in Leiden in September 2008:
**400 Years of Astronomical Telescopes: A Review of History, Science and Technology**

# Overview with links to Nos. 10-13

No. 3.2 – 2011.01.27, update from a version of 2008.05.27:

## Miraculous approval of Hipparcos in 1980: (2)

ABSTRACT: The approval of the Hipparcos mission in 1980 was far from being smooth since very serious hurdles were encountered in the ESA committees. This process is illuminated here by means of documents from the time and by recent correspondence. The evidence leads to conclude that in case the approval would have failed, Hipparcos or a similar scanning astrometry mission would never have been realized, neither in Europe nor anywhere else.
www.astro.ku.dk/~erik/HipApproval.pdf

No. 10 - 2011.03.26:

# Astrometry Lost and Regained

## From a modest experiment in Copenhagen in 1925 to the Hipparcos and Gaia space missions

ABSTRACT: Technological and scientific developments during the past century made a new branch of astronomy flourish, i.e. astrophysics, and resulted in our present deep understanding of the whole Universe. But this brought astrometry almost to extinction because it was considered to be dull and old-fashioned, especially by young astronomers. Astrometry is the much older branch of astronomy which performs accurate measurements of positions, motions and distances of stars and other celestial bodies. Astrometric data are of great scientific and practical importance for investigation of celestial phenomena and also for control of telescopes and satellites and for monitoring of Earth rotation. Our main subject is the development during the $20^{th}$ century which finally made astrometry flourish as an integral part of astronomy through the success of the Hipparcos astrometric satellite, soon to be followed by the even more powerful Gaia mission.
www.astro.ku.dk/~erik/AstromRega3.pdf

No. 11 - 2011.04.06:

# Roemer and Gaia

ABSTRACT: During the Hipparcos mission in September 1992, I presented a concept for using direct imaging on CCDs in scanning mode in a new and very powerful astrometric satellite, Roemer. The Roemer concept with larger aperture telescopes for higher accuracy was developed by ESA and a mission was approved in 2000, expected to be a million times better than Hipparcos. The present name Gaia for the mission reminds of an interferometric option also studied in the period 1993-97, and the evolution of optics and detection in this period is the main subject of the present report. The transition from an interferometric GAIA to a large Roemer was made on 15 January 1998. It will be shown that without the interferometric GAIA option, ESA would hardly have selected astrometry for a Cornerstone study in 1997, and consequently we would not have had the Roemer/Gaia mission.
www.astro.ku.dk/~erik/RoemerGaia.pdf

No. 12 - 2011.01.15:   On the website of the Niels Bohr Institute:

# Surveying the sky

"An astrometric experiment in 1925 was the beginning of a development which Erik Høg, Associate Professor Emeritus, took part in for 50 years. A scientific highlight is the star catalogue Tycho-2 from the year 2000, which describes the positions and movements of 2.5 million stars and is now absolutely essential to controlling satellites and for astronomical observations."

In English:  http://www.nbi.ku.dk/english/www/    and    in Danish:  http://www.nbi.ku.dk/hhh/

and

# En landmåler i himlen

In Danish: En artikel i tidsskriftet KVANT, oktober 2010, om 50 års arbejde

Erindringer om 50 år med astrometrien, der begyndte ved en høstak syd for Holbæk og førte til bygning af to satellitter. Et videnskabeligt højdepunkt er stjernekataloget Tycho-2, der nu er helt uundværligt ved styring af satellitter og ved astronomiske observationer.
www.astro.ku.dk/~erik/kv-2010-3-EH-astrometri.pdf

No. 13 - 2011.03.26:

# Lectures on astrometry

Overview, handout, abstracts at:   www.astro.ku.dk/~erik/Lectures2.pdf

# Brief overview :

**Lecture No. 1.**  45 minutes
 **Astrometry Lost and Regained**
   **From a modest experiment in Copenhagen in 1925**
   **to the Hipparcos and Gaia space missions**
  The lecture has been developed over many years and was held in, e.g., Copenhagen, Vienna, Bonn, Düsseldorf, Vilnius, Oslo, Nikolaev, Poltava, Kiev, Thessaloniki, Ioannina, Athens, Rome, Madrid, Washington, and Charlottesville - since 2007 in PowerPoint.  Revised in 2009 and with the new title

*Astrometry Lost and Regained* it was held in Heidelberg, Sct. Petersburg, Rio de Janeiro, Morelia, Mexico City, Beijing, Montpellier, Groningen, Amsterdam, and Leiden.

---

**Lecture No. 2.**  45 minutes
**Hipparcos - Roemer - Gaia**
   The lectures briefly outlines the development of photoelectric astrometry culminating with the Hipparcos mission. Development of the Gaia mission beginning in 1992 is followed in detail.
   The lecture has been held since 2010 in Toulouse and at ESTEC in Holland.
_________________________________________________________________

**Lecture No. 3.**  45 minutes.  Suited for a broad audience, including non-astronomers
**The Depth of Heavens - Belief and Knowledge during 2500 Years**
   The lecture outlines the structure of the universe and the development of science during 5000 years, focusing on the distances in the universe and their dramatic change in the developing cultural environment from Babylon and ancient Greece to modern Europe.
   The lecture was first held in 2002, and since 2007 in PowerPoint. Held in Copenhagen, Vilnius, Nikolaev, Athens, Catania, Madrid, and Paris
       Handouts at:   www.astro.ku.dk/~erik/DepthHeavens2.pdf
       and    www.astro.ku.dk/~erik/DepthHeavens.pdf

   **An article with the same title as the lecture** appeared in Europhysics News (2004) Vol. 35 No.3. Here slightly updated, 2004.02.20:   www.astro.ku.dk/~erik/Univ7.5.pdf

---

**Lecture No. 4.**  45 or 30 minutes.
**400 Years of Astrometry: From Tycho Brahe to Hipparcos**
   The four centuries of techniques and results are reviewed, from the pre-telescopic era until the use of photoelectric astrometry and space technology in the first astrometry satellite, Hipparcos, launched by ESA in 1989.
   The lecture was presented as invited contribution to the symposium at ESTEC in September 2008: **400 Years of Astronomical Telescopes: A Review of History, Science and Technology.** The report to the proceedings is included as No. 8 among the "Contributions to the history of astrometry ".
   ++++++++++++++++++++++++++++++++++++++++++++++++++++++++++++++++

Further installments in preparation:   On the Hipparcos mission studies 1975-79 and on the Hipparcos archives.

*Best regards     Erik*          *http://www.astro.ku.dk/~erik*

# Reports from 2008 and 2011 on History of Astrometry:

Overview, summary and link to individual reports from 2008 and 2011 are placed in an index file: www.astro.ku.dk/~erik/erik-hoeg-history-of-astrometry-1104-index.pdf .

The two collections of reports are placed in two big files at the following links, including overview and summary pages:

The reports from 2008 are placed at arXiv and in a file printing on 8+94 pages: www.astro.ku.dk/~erik/HistoryAll.pdf    and the title is:
"Astrometry and optics during the past 2000 years"

The reports from 2011 are placed at arXiv and in a file printing on 8+46 pages: www.astro.ku.dk/~erik/History2All.pdf    and the title is:
"Astrometry during the past 100 years"

# Astrometry during the past 100 years

Erik Høg  Niels Bohr Institute, Copenhagen

2011.05.03:

ABSTRACT: Reports from 2008: *Astrometry and optics during the past 2000 years*, are available at www.astro.ku.dk/~erik/History.pdf . - Here are now further *contributions to the history of astrometry related to space astrometry*. The development of photoelectric astrometry is followed from an experiment in 1925 up to the Hipparcos satellite mission in the years 1989-93. This period continues with my proposal in 1992 for CCD astrometry with a scanning satellite called Roemer, which led to the Gaia mission due for launch in 2013. Lectures on astrometry are described. -  Further installments are planned.

The *short* file at www.astro.ku.dk/~erik/History2.pdf of 4 pages contains a table of contents and an overview with links to the individual reports.

The *big* file at www.astro.ku.dk/~erik/History2All.pdf of 11MB will print on 8+46 pages. It contains a table of contents, an overview with links, and all the new reports.

## CONTENTS

# Overview with links

No. 3.2 – 2011.01.27, update from a version of 2008.05.27:

## Miraculous approval of Hipparcos in 1980: (2)

ABSTRACT: The approval of the Hipparcos mission in 1980 was far from being smooth since very serious hurdles were encountered in the ESA committees. This process is illuminated here by means of documents from the time and by recent correspondence. The evidence leads to conclude that in case the approval would have failed, Hipparcos or a similar scanning astrometry mission would never have been realized, neither in Europe nor anywhere else.
www.astro.ku.dk/~erik/HipApproval.pdf

No. 10 - 2011.03.26:

## Astrometry Lost and Regained

### From a modest experiment in Copenhagen in 1925 to the Hipparcos and Gaia space missions

ABSTRACT: Technological and scientific developments during the past century made a new branch of astronomy flourish, i.e. astrophysics, and resulted in our present deep understanding of the whole Universe. But this brought astrometry almost to extinction because it was considered to be dull and old-fashioned, especially by young astronomers. Astrometry is the much older branch of astronomy which performs accurate measurements of positions, motions and distances of stars and other celestial bodies. Astrometric data are of great scientific and practical importance for investigation of celestial phenomena and also for control of telescopes and satellites and for monitoring of Earth rotation. Our main subject is the development during the 20$^{th}$ century which finally made astrometry flourish as an integral part of astronomy through the success of the Hipparcos astrometric satellite, soon to be followed by the even more powerful Gaia mission.
www.astro.ku.dk/~erik/AstromRega3.pdf

No. 11 - 2011.04.06:

## Roemer and Gaia

ABSTRACT: During the Hipparcos mission in September 1992, I presented a concept for using direct imaging on CCDs in scanning mode in a new and very powerful astrometric satellite, Roemer. The Roemer concept with larger aperture telescopes for higher accuracy was developed by ESA and a mission was approved in 2000, expected to be a million times better than Hipparcos. The present name Gaia for the mission reminds of an interferometric option also studied in the period 1993-97, and the evolution of optics and detection in this period is the main subject of the present report. The transition from an interferometric GAIA to a large Roemer was made on 15 January 1998. It will be shown that without the interferometric GAIA option, ESA would hardly have selected astrometry for a Cornerstone study in 1997, and consequently we would not have had the Roemer/Gaia mission.
www.astro.ku.dk/~erik/RoemerGaia.pdf

No. 12 -  2011.01.15:   On the website of the Niels Bohr Institute:

# Surveying the sky

"An astrometric experiment in 1925 was the beginning of a development which Erik Høg, Associate Professor Emeritus, took part in for 50 years. A scientific highlight is the star catalogue Tycho-2 from the year 2000, which describes the positions and movements of 2.5 million stars and is now absolutely essential to controlling satellites and for astronomical observations."

In English:  http://www.nbi.ku.dk/english/www/   and   in Danish:  http://www.nbi.ku.dk/hhh/

and

# En landmåler i himlen

In Danish: En artikel i tidsskriftet KVANT, oktober 2010, om 50 års arbejde

Erindringer om 50 år med astrometrien, der begyndte ved en høstak syd for Holbæk og førte til bygning af to satellitter. Et videnskabeligt højdepunkt er stjernekataloget Tycho-2, der nu er helt uundværligt ved styring af satellitter og ved astronomiske observationer.
www.astro.ku.dk/~erik/kv-2010-3-EH-astrometri.pdf

No. 13 -  2011.03.26:

# Lectures on astrometry

Overview, handout, abstracts at:   www.astro.ku.dk/~erik/Lectures2.pdf

## Brief overview :

**Lecture No. 1.**  45 minutes
 **Astrometry Lost and Regained**
  **From a modest experiment in Copenhagen in 1925**
  **to the Hipparcos and Gaia space missions**
  The lecture has been developed over many years and was held in, e.g., Copenhagen, Vienna, Bonn, Düsseldorf, Vilnius, Oslo, Nikolaev, Poltava, Kiev, Thessaloniki, Ioannina, Athens, Rome, Madrid, Washington, and Charlottesville - since 2007 in PowerPoint.  Revised in 2009 and with the new title *Astrometry Lost and Regained* it was held in Heidelberg, Sct. Petersburg, Rio de Janeiro, Morelia, Mexico City, Beijing, Montpellier, Groningen, Amsterdam, and Leiden.

___

**Lecture No. 2.**  45 minutes
**Hipparcos - Roemer - Gaia**
  **The lectures briefly outlines the development of photoelectric astrometry culminating with the Hipparcos mission. Development of the Gaia mission beginning in 1992 is followed in detail.**
  The lecture has been held since 2010 in Toulouse and at ESTEC in Holland.
_______________________________________________________________________

**Lecture No. 3.**   45 minutes.   Suited for a broad audience, including non-astronomers

**The Depth of Heavens - Belief and Knowledge during 2500 Years**

   **The lecture outlines the structure of the universe and the development of science during 5000 years, focusing on the distances in the universe and their dramatic change in the developing cultural environment from Babylon and ancient Greece to modern Europe.**

   The lecture was first held in 2002, and since 2007 in PowerPoint. Held in Copenhagen, Vilnius, Nikolaev, Athens, Catania, Madrid, and Paris

      Handouts at:   www.astro.ku.dk/~erik/DepthHeavens2.pdf
   and   www.astro.ku.dk/~erik/DepthHeavens.pdf

   **An article with the same title as the lecture** appeared in Europhysics News (2004) Vol. 35 No.3. Here slightly updated, 2004.02.20:   www.astro.ku.dk/~erik/Univ7.5.pdf

---

**Lecture No. 4.**   45 or 30 minutes.

**400 Years of Astrometry: From Tycho Brahe to Hipparcos**

   **The four centuries of techniques and results are reviewed, from the pre-telescopic era until the use of photoelectric astrometry and space technology in the first astrometry satellite, Hipparcos, launched by ESA in 1989.**

   The lecture was presented as invited contribution to the symposium at ESTEC in September 2008: **400 Years of Astronomical Telescopes: A Review of History, Science and Technology.** The report to the proceedings is included as No. 8 among the "Contributions to the history of astrometry ".

++++++++++++++++++++++++++++++++++++++++++++++++++++++++++++++++++

Further installments in preparation:   On the Hipparcos mission studies 1975-79 and on the Hipparcos archives.

*Best regards     Erik*          *http://www.astro.ku.dk/~erik*

2011.01.27

**Miraculous approval of Hipparcos in 1980: (2)[1]**

*Erik Høg, Niels Bohr Institute, Copenhagen*

ABSTRACT: The approval of the Hipparcos mission in 1980 was far from being smooth since very serious hurdles were encountered in the ESA committees. This process is illuminated here by means of documents from the time and by recent correspondence. The evidence leads to conclude that in case the approval would have failed, Hipparcos or a similar scanning astrometry mission would never have been realized, neither in Europe nor anywhere else.

## 1. Introduction

The discussions in ESAs Astronomy Working Group (AWG) and the Science Advisory Committee (SAC) in 1979-80 have been summarised in a previous report (Høg 2008) as repeated here in section 2. I have in the present report chosen to let documents and witnesses speak separately, through quotations and recent correspondence. It may look a bit complicated, but I hope at least some readers will appreciate to get in closer touch with history in this manner.

Correspondence with Ed van den Heuvel is collected in section 3, and I am quoting in extenso because I think the drama is of some interest for a wider audience. Section 4 brings further quotations from the meetings in AWG, SAC, and the Scientific Programme Committee (SPC) and from recent correspondence with Jean Kovalevsky and Catherine Turon. I conclude that Hipparcos prevailed thanks to a kind of miracle. In section 5 I argue that in case the approval would have failed, Hipparcos would never have been realized.

Lennart Lindegren just wrote that he intends to write down the developments up to 1980 from his own perspective, but he cannot promiss a certain date. Jean Kovalevsky will try to write before summer on the 1965-1975 period. I will update the present report if further evidence of sufficient interest should become available.

---

[1] This report is identical to that of 2008-05-28, except that I have added a note in January 2011 at the end of section 3 which shows the crucial role of E.P.J. van den Heuvel in the AWG decision as advocate of Hipparcos. The remaining text and conclusions of 2008 are unchanged.

## 2. Summary of discussions in AWG, SAC, and SPC

The Hipparcos project won the competition with the EXUV project in ESAs Astronomy Working Group, but only barely so according to Edward van den Heuvel (2008, priv. comm.), X-ray astronomer and a member of AWG until the end of 1979, and much in favour of Hipparcos. Several votings took place in AWG before 1980, and at one of the crucial ones Hipparcos stayed for further consideration only because one person had been convinced to change position.

My own attitude then was that if Hipparcos had lost I was ready to quit the project for lack of faith that the astrophysicists would ever let it through.

The final voting in AWG took place on 24 January 1980 (ESA 1980a): Of the 13 members present, 8 voted in favour of Hipparcos and 5 in favour of EXUV, but dangers for Hipparcos laid ahead. At its meeting on 6th and 7th February 1980 the Science Advisory Committee (SAC) discussed six missions and preferred (ESA 1980b) the combined Comet/Geos-3 mission and the Hipparcos mission. The SAC did not make the choice between these two missions which represented the interests of the ESA working groups for respectively the solar system and astronomy. Both missions were therefore recommended, though on certain conditions, and the process ultimately led ESA to do something ESA had never done before: approve two missions at the same time. SAC expressed a preference for Hipparcos over the EXUV mission if the payload is funded *outside* the mandatory budget of ESA. In the end Hipparcos was funded *within* the mandatory budget, so Hipparcos was up against great hurdles all the time, but our mission won in the end, thanks to negotiations of which details are reported by Jean Kovalevsky in section 4. This leads to a summary of the ESA committee meetings in January to July of 1980:

**24 Jan. AWG:** Hipparcos is recommended.

**6/7 Feb. SAC:** Comet/Geos3 and Hipparcos are recommended, no choice is made within SAC, but there are conditions on both.

**4/5 Mar. SPC:** Hipparcos is selected as the next scientific project of ESA. The Hipparcos instrumental payload is included on certain conditions. The mission to Halley comet shall be pursued on certain conditions, and if these conditions are met SPC will in fact have approved two missions simultaneously, resulting in consequences for the schedules.

**8/9 July SPC:** Giotto is included for a flyby in 1986 of Comet

Halley as a purely European project since NASA could not make a firm commitment. The schedule of Hipparcos is accordingly stretched by six months.

## 3. Edward van den Heuvel (2008 and 2011, priv. comm.)

The summary in the first paragraph of section 2 was based on the following mails, here slightly shortened and quoted with permission from Ed van den Heuvel. I asked Ed on 17 March 2008 how close the vote in AWG was. He answered at 6:07 PM our time, the same day:

Dear Erik,
The vote was indeed very close. I was able to convince one of the X-ray astronomers (Spada) not to vote for the EUV/Soft X-ray mission which was then the competitor of Hipparcos, and his vote was just the one that made the difference ....

Spada, although director of the X-ray astronomy lab in Bologna, casted the vote that made the difference

very sadly, Spada has completely disappeared from the scene in Italy. …

I am at the moment working at the Institute for Theoretical Physics, University of California Santa Barbara. If you wish to call me  …

Best wishes,
Ed van den Heuvel

**An hour later, at 7:21 he added:**

Dear Erik,

It is a long time ago, and there have perhaps been various stages of voting in the AWG. I do not have any of my papers here in California, so I cannot check.I know I kept my papers from that time in the AWG in my archive in Amsterdam, so when I am back I can check.

What I remember is that we first had Setti as the AWG chair (I thought you were in the AWG at that time), and under his chairmanship we had many discussions of the projects but not a final vote. When the vote had to be taken, Setti had been replaced by De Jager from my country, who had a big stake in the EUV/X-ray mission. .... It was under his guidance that the vote which I mentioned in my last e-mail to you was taken and in which Spada and I (as X-ray astronomers) voted in favour of Hipparcos ...

Now that you say that I was no longer in the AWG in 1980 when apparently a final vote was taken, I am getting a bit confused, about whether there may have been a still later (definitive?) round of votes and whether the votes which I mentioned was perhaps an earlier round.

I presume that it must be possible to trace that back in the minutes of the AWG from 1979 and 1980.

As you know, memory is not fully reliable, and this was almost 30 years ago. But I vividly remember that there was this one voting round where Spada's vote made the difference. I thought that what I remembered is that if in that voting round Hipparcos would have lost, then the AWG from that moment would have gone further with the EUV/X mission. But I hope this can be traced back in the AWG minutes.

There you also could trace back whether Spada was still in the AWG when the final vote was made. I do not know whether the minutes tell whom voted in favour and whom voted against? *(No, the minutes do not give such details, EH)*

Since I am just saying this all from the top of my head, without any papers here that may support it, and since- as said- memory may be unreliable, please consider all this as confidential, and not for circulation. *(Permission has later been given, EH)*

Best wishes,
Ed

*Note by EH: It seems that Ed has been member of AWG with his period of three years 1976-79 overlapping my years 1976-78. But I do not remember him from that time in spite of his great sympathy for the space astrometry project and the important role he has played in the mission approval. About twenty years ago, however, he told me what I just reported, and he has recalled it ever since when we happened to meet with years between. Therefore I contacted him when I was writing (Høg 2008) and got immediate reply.*

***Note by EH added in January 2011 with Ed's permission:*** *A conversation in Amsterdam with Ed resolved the questions of doubt mentioned above by Ed. The round of vote in AWG mentioned was in fact the final one on 24 January 1980 where the X-ray astronomer Spada voted for Hipparcos which would otherwise have lost to the EXUV mission. Also radio astronomer Schilizzi voted in favour, after consulting with Ed. This gave the vote of 8 to 5 in favour of Hipparcos. Present at the meeting as members of AWG were thirteen persons: de Jager, Cezarsky, Delache, Drapatz, Fabian, Grewing, Jamar, Murray, Perola, Puget, Schilizzi, Spada, and Swanenburg while Rego was unable to attend.*

*Van den Heuvel, although no longer a member of AWG, and Delache had the preceding day on invitation by the chairman, de Jager, presented a summary of the two missions "on behalf of the Chairman ... to assist the Working Group in its formulation of the recommendation" (quoted from the letter of invitation). It was quite unexpected by de Jager who was Ed's former boss and also an X-ray astronomer, and not to his liking that Ed strongly advocated Hipparcos.*

## 4. From the committee meetings in 1980

Some further quotations from AWG and SAC meetings (ESA 1980a and 1980b) illustrate the difficulties

Hipparcos encountered. At a meeting on 24 January 1980 the AWG considered the Astrometry and EXUV missions, concluding that both missions will give excellent scientific return. This is elaborated for the two missions. On astrometry for instance this: *"The Astrometry mission, HIPPARCOS, will give fundamental quantitative results to all branches of Astronomy. It emphasises typical European know how and will serve a community never before involved in space research"*; on the EXUV mission for instance this: *"The fact that the scientific objectives of this mission are being covered by two different missions proposed by other agencies (EUVE by NASA and ROBISAT by Germany) emphasises its timeliness."*

It is somewhat surprising then that 5 members were still in favour of EXUV and only 8 in favour of HIPPARCOS. One could have thought that a unique mission as Hipparcos would come above anything else in everybody's mind.

SAC discussed the missions on 6$^{th}$ and 7$^{th}$ February 1980 and unanimously recommended that the combined Comet/Geos-3 mission be selected as proposed by the Solar System Working Group (SSWG) on certain conditions. Strong advocates for EXUV were also present at the SAC meeting: *"in the event that the Hipparcos payload would need to be funded within the mandatory programme, the SAC was divided as to whether Hipparcos should then remain the Agency's choice or <u>EXUV</u> should be carried out because this mission was considered by some members to be just as interesting."* (The quotation is literal, including spellings and emphasis.) In the end, Hipparcos was in fact financed *within* the mandatory programme.

In view of all these hurdles it seems a kind of miracle that Hipparcos could prevail, but it was of course because the right people worked hard to make it happen. The final solution was that SPC approved two missions: Giotto, the mission to comet Halley, to be launched first and to be followed by Hipparcos, and that SPC decided to finance the Hipparcos scientific payload out of the mandatory programme. ESA otherwise always assumes that payloads are financed by the member states.

Where were the competing EXUV people in all this? An answer may be found in the following letters from Jean Kovalevsky.

**Jean Kovalevsky** *wrote on 2008.05.11:*
*I was invited to the AWG for the Hipparcos presentation, but did not attend the discussions.*

*I was member of SAC and I remember very well that, at some point, there was a vote between Hipparcos and EXUV: Hipparcos had 5 votes out of 6, the only tenant of EXUV was H Elliot from the UK. The other members were: Egidi (Frascati), Tammann (Basel), Weiss (Erlangen) and Pinkau (Chairman). The fact that SAC proposed that Hipparcos payload was to be paid nationally was simply repeating the SSWG statement.*

*It was evident for me and (at least as far as I remember) Tammann, that the responsibility of the payload had to be taken over by ESA, but I felt that insisting on this point would have been counter-productive, because the announced costs of the two proposals without the payload were identical while adding 50 MAU to the cost of Hipparcos would have killed it.*

*So I decided, in order to save the mission, to accept this point. After all, SAC was only an advisory group and had no financial responsibility. The only ESA body that could overrule the normal procedure (following which nations should fund and prepare the payload) was the SPC. An additional problem was that the laboratories involved in space hardware had experience in receivers and in conventional optics, but no one was reasonably able to built the delicate parts of Hipparcos. I knew that at least the French delegation at SPC, and possibly others will lobby in favour of an indoor payload funding. The March decision by SPC proved that I was right.*

*Pinkau had reported to the March SPC meeting of the views of SAC. I prepared, as an attachment for you, the part which concerns Hipparcos and EXUV.*

*From the part on Hipparcos: "The SAC realized the extremely fundamental nature of the mission, and the impact it will have on many branches of science and our conception of the world we live in. The SAC also noted the strong support for this mission within the AWG." Then the three areas of concern to the SAC are outlined: Technical difficulties, the data analysis problem, and the cost of the mission.*

**Catherine Turon** *wrote on 2008.05.13:*
*Hipparcos was approved in March 1980, and Giotto later, after still another meeting of the SPC (exceptional ???), in July 1980. I do not have the minutes of these SPCs neither their decisions, but the letter of information sent to "the wide scientific community" by E.A. Tredelenburg, then Director of the Scientific Programme. I'll send these to you.*

**EH** *wrote on 2008.05.15:*
*I was the only astrometrist in the AWG about 1977 and I remember saying to Malcolm Longair in a coffee break: "You astrophysicists will decide about the astrometry project and you should be aware that you have only one opportunity to approve such a mission. It you reject it this time it cannot be revived because the astrometrists would never again believe astrophysicists could ever let it pass. We would believe that no matter how much you are impressed by space astrometry, in the end the majority would always put their own project higher." He said that I should not use this as an argument, but only*

*argue with the qualities of the project. That was all he said, a wise advice, I think, which I followed. But the insight I believed to have then has become certainty after seeing the evidence presented here.*

**Jean Kovalevsky** *wrote on 2008.05.23:*

*Dear Erik.*

*Let me make some further remarks that could enrich your text, a text which I fully appreciate.*

*Coming back to the February 1980 SAC meeting, there was really NO competition between the Comet/Geos3 mission and the astronomical missions. From the very beginning of its session, SAC did not like the idea of choosing between an astronomical and a Solar system mission. It considered that it would be more fair to give a chance to both working groups' proposals, and that ESA, rather than deciding missions one by one every year or so, must have a broader and more prospective policy.*

*So, indeed, the choice was only between EXUV and Hipparcos. I think that the key sentence in the pages I sent you is the following:"It was thought that then a new proposal for an EUV-mission would be very worthwhile". This was really killing EXUV.*

*Now, there were two conditions:*
*-For Hipparcos, it was the funding of the payload*
*-For the Comet/Geos3 mission, it was the necessary re-assesment to transform it into a really cometary mission.*

*In March, SPC solved the first problem (and this is probably the most miraculous part of the adventure) and, letting time for the re-assesment of the cometary mission, Hipparcos found itself as the ONLY approved mission!*

*What followed is interesting. The re-assesment of the cometary mission, becoming Giotto, put ESA in an awkward situation: the non-approved mission was evidently more urgent because of Halley's orbit. We had an additional SAC meeting end of June or July. I do not have documentation on it, but I remember well how insistently Trendelenbourg (Director of Science) tried to convince me (as he assumed I was the toughest proponent of Hipparcos), that I should accept that Hipparcos be delayed by a year or so, to allow the maximum money to be spent on Giotto. Of course, SAC unanimously agreed and the next SPC followed the recommendation.*

*The decision of the SPC that the payload should be the responsibility of ESA was taken very seriously and ESA started to study how to manage it. In the October 1980 meeting of SAC, the Executive presented a document which described the management as we have known it, and SAC approved it.*

*Best regards,*

*Jean*

**Catherine Turon** *agreed to this later the same day, and did not want to add anything.*

**EH** *wrote on 2008.05.26:*

*The reports mentioned by Catherine have been received (ESA 1980c and d). They spell out in detail what Jean has said in his two letters. Finally, therefore, the summary of the ESA committee meetings in January to July of 1980 can be written and is placed at the end of section 2.*

**5. In case the approval had failed**

It appears that the approval could well have failed in which case I am sure Hipparcos would never have been realized. This proposition has been countered by a colleague:*"You can never know that, something could have happened."* But please consider the situation of astrometry at that time. For decades up to 1980 the astrometry community was becoming ever weaker, the older generation retired and very few young scientists entered the field. I myself would have lost the faith that the astrophysicists would ever let such a mission through, and others would also have left the field of space astrometry.

If someone would have tried a Hipparcos revival one or two decades later the available astrometric competence would have been weaker, and where should the faith in space astrometry have come from? When Hipparcos became a European project in 1975 and the hopes were high for a realization, the competence from many European countries gathered and eventually was able to carry the mission. This could not have been repeated after a rejection of the mission.

But NASA could have realized a Hipparcos-like mission? No, for two reasons: The American astrometric community had much less resources of competence to draw from than there were in Europe, and secondly, as an American colleague said: *"You can convince a US Congressman that it is important to find life on other planets, but not that it is important to measure a hundred thousand stars."*

Thanks to the completion of the Hipparcos mission a strong astrometric community now exists in Europe which has been able to propose and develop the Gaia mission and which will carry it to a successful completion. Without Hipparcos the faith in the much more difficult CCD technology of Gaia would have been missing.

**Acknowledgements:** I am grateful to Catherine Turon for providing the reports ESA 1980a-d, to Edward van den Heuvel for permitting his letters to be included here, and to Jean Kovalevsky for providing more information from the ESA meetings. I also thank all of them and Holger Pedersen for comments to earlier versions of this report.

2011.03.26

# Astrometry Lost[1] and Regained

## From a modest experiment in Copenhagen in 1925 to the Hipparcos and Gaia space missions

*Erik Høg     Niels Bohr Institute*
*Copenhagen University, Denmark*

ABSTRACT: Technological and scientific developments during the past century made a new branch of astronomy flourish, i.e. astrophysics, and resulted in our present deep understanding of the whole Universe. But this brought astrometry almost to extinction because it was considered to be dull and old-fashioned, especially by young astronomers. Astrometry is the much older branch of astronomy, in fact 2000 years of age, which performs accurate measurements of positions, motions and distances of stars and other celestial bodies. Astrometric data are of great scientific and practical importance for investigation of celestial phenomena and also for control of telescopes and satellites and for monitoring of Earth rotation. Our main subject is the development during the $20^{th}$ century which finally made astrometry flourish as an integral part of astronomy through the success of the Hipparcos astrometric satellite, soon to be followed by the even more powerful Gaia mission.

## Synopsis

The renewal of astrometry in the $20^{th}$ century resulted from work by astronomers who saw the necessity and were able to utilize the newest technology for astrometry, and it is now possible in retrospect to see in detail how this was accomplished. The renewal began with a rather modest experiment with photoelectric techniques on the old meridian circle in Copenhagen in 1925 and culminated with the Hipparcos and Gaia space missions. From 1925 to 1975 the ground was laid by a very small number of researchers who, in fact, acted in a single chain: If any of them had been missing we would not have had any Hipparcos mission, and consequently no Gaia.

In 1925 Bengt Strömgren in Copenhagen, Denmark, made experiments with recording of star transits at a meridian circle. He placed a plate with slits in the focal plane with a photocell behind, and recorded the current as a star was moving across. The present author was Strömgren's student 1950 to 1956. Shortly later I went to Hamburg in Germany where I stayed for 15 years. In 1960 I proposed that the recording of star transits should be done with the novel technique of photon counting which was then implemented and used for many years on the Hamburg meridian circle.

In 1967 Pierre Lacroute in Strasbourg, France, proposed to scan the sky with a rotating satellite, recording the star transits with photon counting. Ideas of space astrometry were much studied in France, but only in France. Especially on French initiative, the European Space Agency began studies in 1975 where I was invited to participate. Swiftly, I made a realistic design of a scanning satellite with many new features. The design was studied by astronomers, ESA engineers and industry, and the Hipparcos astrometric mission was approved in 1980. The satellite was launched in 1989 and completed a three year successful mission. The results were published in 1997 and have since been utilized in many thousand publications. Hipparcos observations were obtained with photoelectric detectors, viz. an image dissector tube and two photomultipliers.

In 1992 I proposed a new astrometric mission where CCD detectors were introduced, resulting in a million times higher observing efficiency than Hipparcos achieved. Such a mission, named Gaia, was approved by ESA in 2000 after deep studies by scientists and engineers, and is due for launch in 2013. Thus, astrometry seemed lost, but has been regained through the application of space techniques after astrometric developments during half a century depending critically on a very few men, and subsequently being implemented by large teams of dedicated scientist and engineers.

---

[1] Contribution to the history of astrometry No. 10

Expanded version - the previous was dated 20.02.2010.

## Astrometry seemed lost

The revival of astrometry during the last century was possible through photoelectric astrometry applied to space techniques, implemented in the Hipparcos satellite launched by ESA in 1989. The chain of ideas and experiments which led to Hipparcos is traced in the following, for greater detail see Høg (2008 and 2009).

Photoelectric techniques were used for astrometry by many scientists in the previous century, and the following is not a history of photoelectric astrometry in total. It is limited to the activities which led to Hipparcos. This work was done primarily in Copenhagen, Hamburg, Strasbourg and other places in France, leading from a first experiment on the meridian circle in Copenhagen in 1925 up to approval of the Hipparcos mission by ESA in 1980.

The prospects for astrometry looked bleak at the middle of the 20$^{th}$ century. If an astrometrist retired, the vacancy was usually filled with an astrophysicist, and astrophysics was moving towards the exciting new extragalactic astronomy. But the present author did not feel any pressure from this trend when I studied in Copenhagen (1950-56). My teachers at the observatory, Bengt Strömgren and Peter Naur, were both very familiar with astrometry, and it was natural to follow their advice. As a boy, I had read about Tycho Brahe and Ole Rømer, the two Danish heroes in astronomy, who both worked on what is now called astrometry, astronomy of positions.

In fact, important developments were going on also during the middle of the century which eventually allowed me to lead the construction of the Tycho-2 Catalogue with 2.5 million stars. This catalogue has replaced all previous reference catalogues with its positions and proper motions derived from observations with the Hipparcos satellite and 100 years of ground-based observations. Since its release in 2000, Tycho-2 is being used everywhere to guide astronomical telescopes on the ground and satellites in space, and for astrophysical studies by means of its two-colour photometry.

The term astrometry does not apply to astronomical measurement in general as the word suggests, but only to the measurement of positions on the sky of stars and other celestial objects. The position of a star changes with time due to its proper motion, to the parallactic motion created by the motion of the Earth around the Sun, and to the orbital motion in the case of a binary star.

The term astrometry came into use to distinguish it from astrophysics, especially after the introduction of stellar spectroscopy 150 years ago and of atomic theory later on, which were used to analyse the spectra. For the two millennia prior to that, astrometry had in fact been the main task of astronomy. Astrometric observational data have been the basis for navigation, time keeping and monitoring of Earth rotation, and they have given us a deep astronomical understanding of stars and their distances and motions, star systems, planetary motions, and the underlying physical laws.

The photoelectric effect was discovered in 1887 by Wilhelm Hallwachs. He saw that a negative charge on a zinc plate was lost when it was illuminated by light of sufficiently high frequency, i.e. high energy. The effect was explained in 1905 by Albert Einstein in terms of atomic theory which earned him the Nobel prize in 1921.

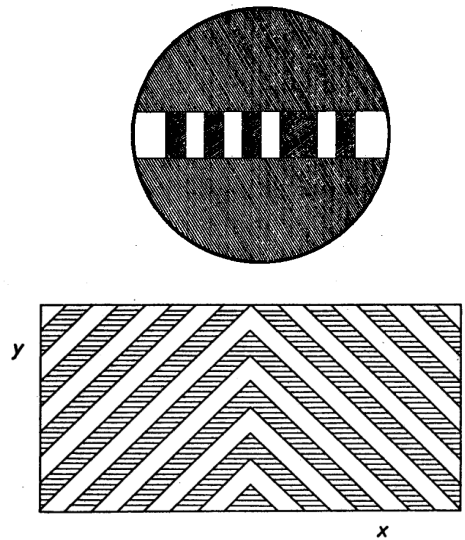

**Fig. 1.** Two slit systems for photoelectric observation with a meridian circle: Strömgren 1925 and Høg 1960, the latter for two-dimensional measurement. In France it was called "une grille de Høg" in the early 1960s.

## Photoelectric astrometry in Copenhagen and Hamburg

Bengt Strömgren was introduced to astronomy by his father who was professor at the Copenhagen University and director of the Observatory. In 1925, at the age of 17 years, he reported about experiments with photoelectric recording of star transits (Strömgren 1925 and 1926). In the focal plane of the old meridian circle in Copenhagen he had placed a system of slits parallel to the meridian, Fig.1. Behind the slits a photocell received the light from the star after it had passed the slits. As the star moved across the slits the variations of light intensity gave corresponding variations in the photo current, and these variations of current were amplified and recorded.

Strömgren, however, found a serious drawback of his initial method: For reasons of statistical noise, it would only allow recording of stars to $6^{th}$ or $7^{th}$ magnitude with a medium size meridian circle. In Strömgren 1933 he therefore proposed a method of integration with a switching mirror and two photocells behind the grid which should allow observation of much fainter stars. But the method posed technical problems and no further experiments have been reported. The present author heard about the two proposals as a student and that bore fruit later on.

In 1940 Bengt Strömgren became director of the observatory which was located in the centre of Copenhagen. The same year he took the initiative to build a new observatory on a hill at the village Brorfelde 50 km west of Copenhagen. The main instrument, a new meridian circle, was installed in 1953 and I got the task as student to test the stability of the new instrument by photographic observations of a star very close to the North Pole.

Most important for me as a young scientist, was to grow up in an environment where a new meridian circle was the main instrument and where this course for the institute had been defined by an outstanding scientist. Bengt Strömgren gave everybody, not only a youngster as me, confidence about the future line of astronomy. How very different at most other places in the world where astrometry, the astronomy of positions, was being discarded as old-fashioned science. At such places I would probably have become an astrophysicist, since I certainly did not want to do old stuff.

My studies finished, I became a conscript soldier. Most of the time I had the opportunity to work in a laboratory measuring radioactive decay of dust, collected to follow the nuclear weapon testing of the two superpowers. This involved radioactive counting techniques and my experience with this brand new technique was later applied to photoelectric astrometry.

In 1958 I moved to the Hamburg Observatory where both astrometry and astrophysics were held in high esteem; Otto Heckmann was the powerful director. I wanted to classify stars by objective prism spectra obtained with the big Schmidt telescope and I built a punched card recording system for the spectrum scanner, something new for that time. But in 1960 I returned to astrometry after the excursion in direction of astrophysics and stellar astronomy. I had the idea (Høg 1960) that Strömgren's method with the switching mirror could be implemented very elegantly by a photon counting technique which I had learnt from the counting of radioactive decay.

A photo multiplier tube should be placed behind a slit system and the photo-electrons be counted in short time intervals, controlled by an accurate clock, and the counts be recorded on punched tape. Later numerical analysis of the counts in a computer would give the transit times across the slits. In principle, the transit time for individual slits could be derived, or the transit time for a group of slits. The latter method would be less sensitive to noise, and in the course of time both methods have been widely applied.

The slits should be inclined to the stellar motion by 45 degrees in alternating directions, Fig. 1. By such a "fishbone grid" a two-dimensional measurement of the star in the focal plane became possible, corresponding to right ascension and declination.

Heckmann was immediately interested and the method was implemented on the Hamburg meridian circle for the expedition to Perth, Western Australia. That kept me busy for the next decade and resulted in a catalogue in 1976 with positions of 25,000 stars.

Astrometry by means of accurate slits and photon counting was subsequently applied on meridian circles, on long-focus telescopes, and ultimately on the first astrometric satellite, Hipparcos. French astronomers became interested in the method, and there were reports from Lille and Besançon in the early 1960s and later followed, Sauzeat (1974) and Creze et al. (1982) where they worked with "une grille de Høg", as they called the system of inclined slits. The method with the fishbone grid and photon counting was crucial in the proposal for space astrometry by Pierre Lacroute.

## Astrometry with a scanning satellite

Pierre Lacroute, director of the Strasbourg Observatory, presented a project of space astrometry at the General Assembly of the International Astronomical Union (IAU) in Prague in 1967 (Lacroute 1967). Lacroute had already presented such a project in a meeting in Bordeaux on 4-6 October 1965, in front of French and Belgian astronomers. This was the first time that such type of astronomy was proposed for a space mission.

The potential advantages were clear, no atmosphere and no gravity, and perhaps thermal stability if that would be technically feasible. I attended the presentation in Prague, but to me and most others the technical problems seemed utterly underestimated. The proposal did not start any activity outside France, but Lacroute's great vision was fortunately shared by other French astronomers, especially by Pierre Bacchus and they worked closely together. Also Jean Kovalevsky supported the project and he has recently given an account of the early years (Kovalevsky 2009). He finally had it converted from being a national project to become European, through ESA.

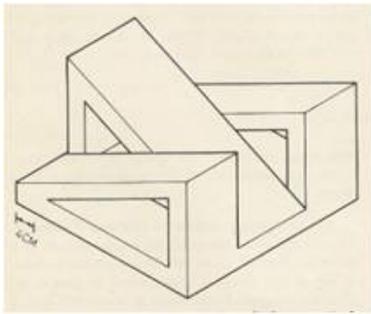

**Fig. 2.** A beam combiner as proposed by Lacroute. Placed in front of the telescope aperture it will combine the beams from two fields on the sky separated by an angle of 90 degrees. The angle will be very stable as defined by the rigid material.

The paper by Lacroute 1974 contains two proposals, a Spacelab option and a free flyer. The Spacelab option requires a telescope with a beam combiner of 40cm x 40cm aperture as in Figure 2 to be flown on 8 missions within two years to observe 40.000 stars. Lacroute's proposal for a free flying scanning satellite is shown in the Figures 2-4, copied from Lacroute 1974.

A study group of astronomers and ESA engineers was set up in 1975 and I joined the group on invitation, in spite of my profound scepticism and lack of interest in space techniques. But the first meeting on 14 October changed my scepticism because the chairman of the meeting urged us not especially to consider the existing proposals, but simply to think about how we could make use of space techniques for our science, astrometry.

That made me think freely, in fact converted me to become an enthusiast, and with a number of major changes in the following weeks I could swiftly transform the satellite proposed by Lacroute, see Figures 2-4. I sent a proposal six weeks later (Høg 2011a) which was technically simpler and vastly more effective because an image dissector tube replaced the photomultipliers. Other equally important new features in my proposal were: One-dimensional measurement along scan, a beam combiner of two parts – not three parts as Fig. 2, change its angle from 45 degrees, use a modulating grid instead of slits, use active attitude control, make the spin axis revolve around the Sun at a constant angle, use a star mapper with one photomultiplier to detect reference stars, use an input catalogue with 100 000 selected stars. All these ideas formed a self-consistent instrument which by mid 1976 looked as Fig. 5, see Høg (1997).

It is interesting to note that the new design in 1975 was based on technology which had been available also e.g. ten years earlier if somebody would have thought of combining it to an astrometric mission. In particular, the detection was made by an image dissector tube instead of photo multiplier tubes which increased the detection efficiency by a factor of one hundred, and the image dissector was developed in the 1930s and had since been widely used as electronic television camera.

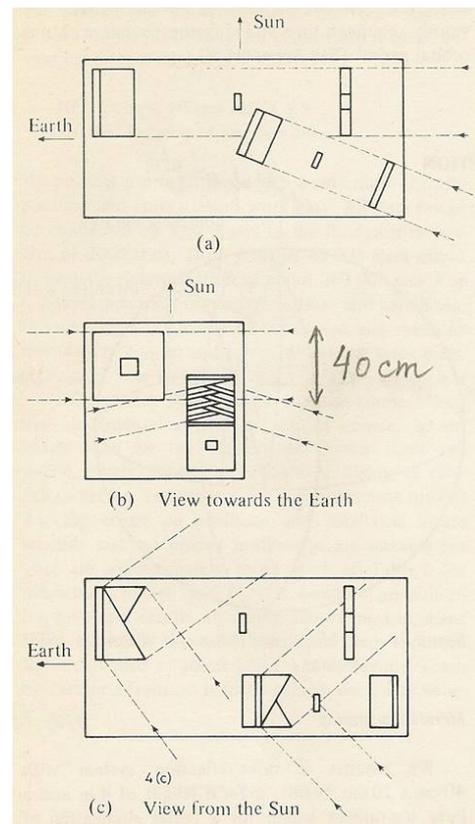

**Fig. 3.** Two telescopes as proposed by Lacroute. By rotation about a spin axis pointing in the direction to the Sun the telescopes will continuously scan the sky with slits as in Fig. 4.

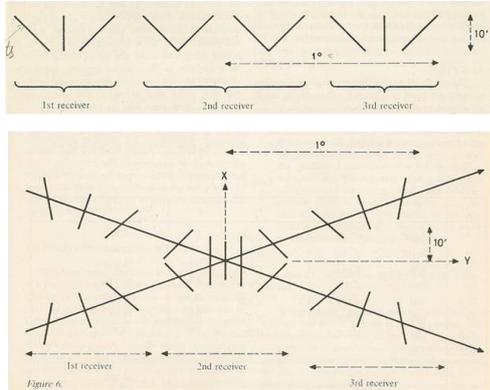

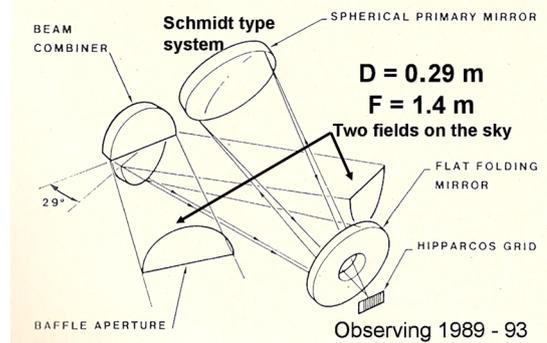

**Fig. 4.** The stars will cross the slits and be measured by six photomultipliers. The upper slit system is used in the larger telescope of Fig. 3, the lower one in the smaller telescope. Since the latter telescope is scanning a small circle on the sky the stars from the two fields move in different directions.

**Fig. 6.** Hipparcos as launched in 1989: Schmidt system with 29 cm diameter aperture, 1.4 m focal length and two viewing directions, all mirrors silver coated for maximum reflectivity. The satellite rotation makes the stars cross the modulating grid and the Tycho star mapper slits.

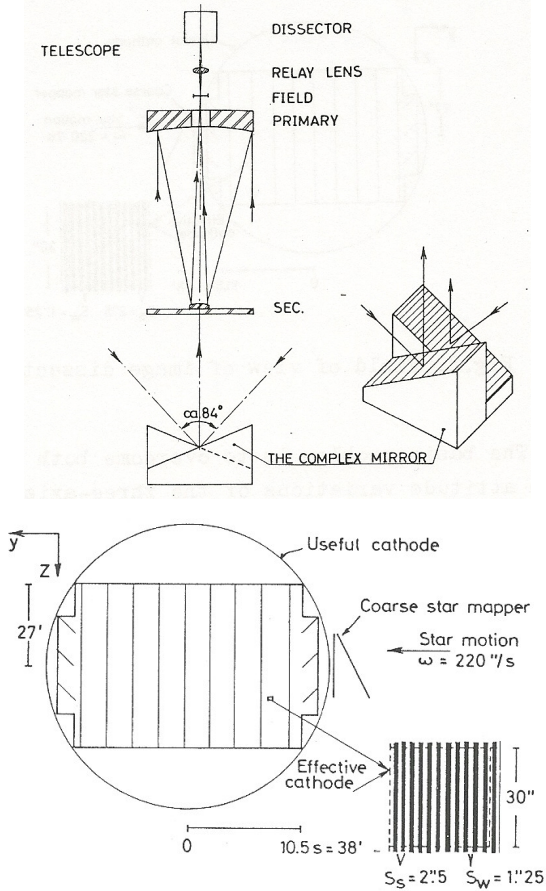

**Fig. 5.** Hipparcos design by mid 1976. When launched in 1989, the telescope was very different, a folded Schmidt system as shown in Fig. 6, but the slit system and detectors were quite similar.

In 1976 the data reduction was a formidable task: to derive positions, proper motions and parallaxes for 100,000 stars from 10 million angular measures. Fortunately, I was already acquainted with Lennart Lindegren since 1973 when he was a 23 year old student at Lund Observatory. On 22 September 1976 I introduced him to Hipparcos and after four weeks he presented the mathematical formulation of the method which was later used during the mission, the "three-step method." Two weeks later came a report with the first simulations. Without his unfailing genius in all mathematical, computational and optical matters the project would not have been ripe for approval in 1980, and probably never.

By the end of 1979, after studies involving astronomers, ESA engineers and the industry, Hipparcos looked very different from the early ideas laid down by Lacroute. His idea of a satellite scanning the sky with a beam combiner mirror viewing in two directions with one telescope was maintained, but it was yet bolder in its objectives and technically more realistic. As a result, it had also succeeded in generating a substantial scientific following across Europe, backed by an increasingly vocal international community; these sentences are partly quoted from a book about the project by Michael Perryman published in 2010.

During the first months of 1980, decision about the next ESA mission was taken in difficult negotiations where an EXUV project and a mission to comet Halley were very strong competitors. The competition ultimately led ESA to do two things the agency had never done before: firstly to approve two missions at the same time, Hipparcos and the Giotto mission to comet Halley, and secondly to finance the Hipparcos payload out of the science budget. Otherwise ESA always paid spacecraft and launch and the national institutes built and financed their experiments to go on board. Hipparcos was up against great hurdles all the time, but our mission won in the end, thanks to negotiations in which Jean Kovalevsky took part. My own attitude then was that if Hipparcos had lost I was ready to quit the project for lack of faith that the astrophysicists would ever let it through.

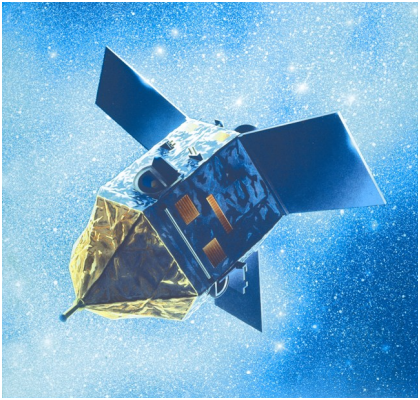

**Fig. 7.** The Hipparcos satellite launched 1989.

In April 1981 the satellite was well into the design phase when significant modifications would normally have been strictly rejected for reasons of risk, and for the increased cost that they would incur. But at that time I realised that the signals from the satellite attitude detectors, i.e. the star mapper slits, contained an enormous quantity of star positions that were not being sent to the ground. I immediately pointed out in three reports to ESA what was at stake and the modifications to the design were made, including the addition of colour filters and detectors.

This "Tycho experiment" as it was called, resulted in the Tycho-2 Catalogue in 2000 with astrometry and two-colour photometry of 2.5 million stars (Høg et al. 2000). Tycho-2 is now the preferred astrometric reference catalogue for star brighter than 11$^{th}$ magnitude, used to tie the bright 120,000 stars of the Hipparcos system to astrometric observations of fainter stars obtained by ground-based CCD telescopes.

After approval the project gained great momentum and was carried through by large enthusiastic teams (Perryman et al. 1997) working many years guided by the Hipparcos Science Team whose chairman Michael Perryman personifies this phase of the mission more than anyone.

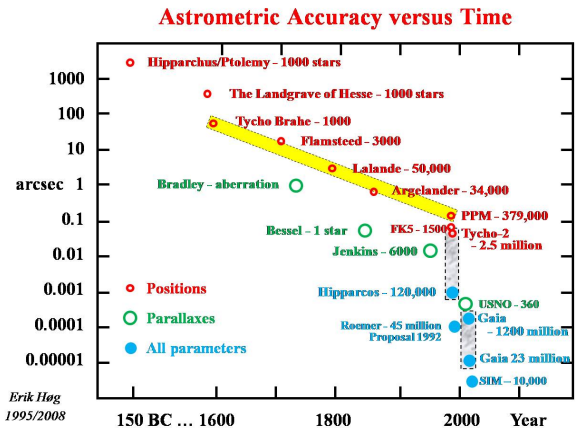

**Fig. 8.** Astrometric accuracy during the past 2000 years. The accuracy was greatly improved shortly before 1600 by Tycho Brahe. The following 400 years brought even larger but much more gradual improvement before space techniques with the Hipparcos satellite started a new era of astrometry.

## Astrometry regained!

Hipparcos was launched in 1989, Figures 6 and 7, observed for three years, and the results were extensively published in 1997. The two cited papers about the Hipparcos and Tycho Catalogues by Perryman et al. (1997) and Høg et al. (2000) are among the 40 most cited articles in Astronomy & Astrophysics out of 50,000 published in 40 years and have therefore been reprinted recently in Volume 500. The Hipparcos Catalogue of 1997 has been superseded by a new reduction of all raw observation data by van Leeuwen (2007) resulting in what may be called Hipparcos-2. The bright stars are much more accurate in this catalogue with the result, e.g., that 30,000 stars obtain distances with less than 10 per cent error, compared to 21,000 in the catalogue from 1997, and to less than 1000 stars before Hipparcos. These facts and the Figure 8 illustrate the revival of astrometry.

Bengt Strömgren appears clearly at the root of my contributions to astrometry, including Hipparcos, and he was directly active before the mission approval in 1980 in order to ensure Danish and Swedish support. It seems from the unbroken chain of actions listed above and detailed in Høg (2008 and 2009) that there would have been no Hipparcos, no space astrometry with a scanning satellite, if any of the four persons Bengt Strömgren, Pierre Lacroute, Jean Kovalevsky or Lennart Lindegren had been absent from the scene before 1980, and I may include myself and Otto Heckmann for his immediate strong support of my ideas, thus Fig. 9 with the six astronomers.

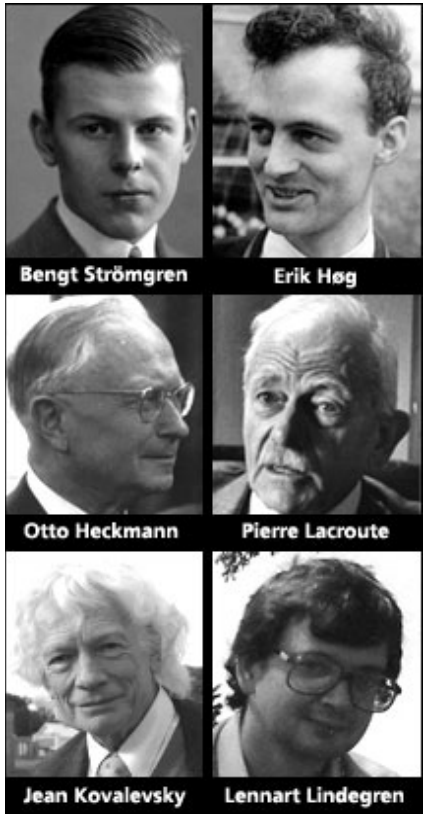

**Fig. 9.** The development of photoelectric astrometry since 1925 and of the Hipparcos project was critically dependent on every one of these six astronomers up to the approval in 1980.

Finally, the crucial role of Edward van den Heuvel in the final decision of AWG (ESA's Astronomy Working Group) on 24 January 1980 as advocate of Hipparcos must be pointed out. Without van den Heuvel, Hipparcos would have lost to the EXUV mission (EXtreme UltraViolet) and nothing could have changed that decision. Many had worked for the development of photoelectric astrometry and of Hipparcos and for a positive decision in 1980, but seven persons virtually formed a chain in which every link was indispensable. The whole ESA decision process has been described in Høg (2011b) which has recently been updated.

It appears that the approval by ESA could well have failed, in which case I am sure Hipparcos would never have been realized. This proposition has been countered by a colleague: *"You can never know that, something could have happened."* But please consider the situation of astrometry at that time. For decades up to 1980 the astrometry community was becoming ever weaker, the older generation retired and very few young scientists entered the field. I myself would have lost the faith that the astrophysicists would ever let such a space mission through, and others would also have left the field of space astrometry.

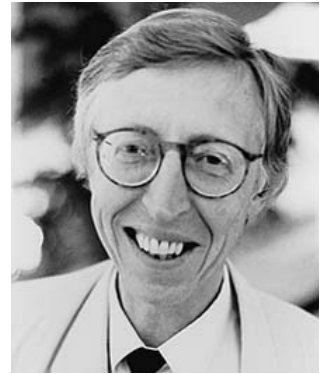

**Fig. 10.** Edward van den Heuvel strongly advocated Hipparcos in the ESA decision in 1980 although he himself as an X-ray astronomer had a direct interest in the EXUV mission.

If someone would have tried a revival of the idea one or two decades later, the available astrometric competence would have been weaker, and where should the faith in space astrometry have come from? When Hipparcos became a European project in 1975 and the hopes were high for a realization, the competence from many European countries gathered and eventually was able to carry the mission. This could not have been repeated after a rejection of the mission.

But could NASA have realized a Hipparcos-like mission? No, and for two reasons: The American astrometric community had much less resources of competence to draw from than available in Europe, and secondly, as an American colleague said: *"You can convince a US Congressman that it is important to find life on other planets, but not that it is important to measure a hundred thousand stars."*

Thanks to the completion of the Hipparcos mission a strong astrometric community now exists in Europe which has been able to propose and develop the Gaia mission, Figure 11, and which will carry it to a successful completion. Without Hipparcos the faith in the much more difficult CCD technology of Gaia would have been missing. Gaia is a scanning satellite with two directions of view imaged directly on a focal plane, similar to Hipparcos. The star images are however measured not by a photoelectric image dissector tube as in Hipparcos but by a large mosaic of CCDs as proposed in 1992 for the Roemer mission (Høg 1993, 2007). The name Gaia still reminds of an option GAIA, where the "I" stood for Interferometry, which was intensely studied 1993-97, but then abandoned in January 1998. The development continued with the Roemer concepts and large aperture telescopes – Gaia is a large Roemer, see Høg (2011c).

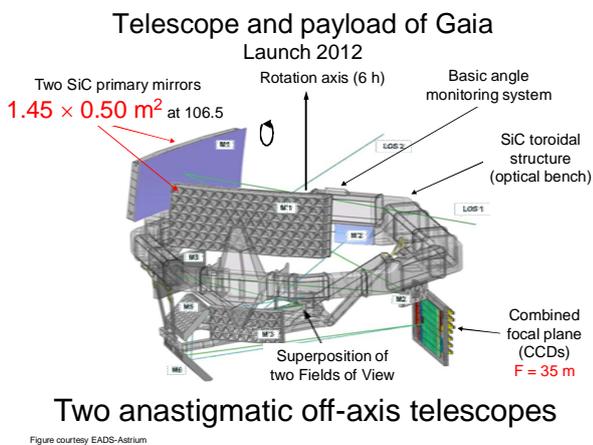

**Fig. 11.** The Gaia payload, final design of 2005.

**Acknowledgements**  Comments to previous versions of this paper from J. Kovalevsky, H. Pedersen, M. Perryman, C. Turon, and F. van Leeuwen are gratefully acknowledged.

2011.04.06

# Roemer and Gaia[1]

Erik Høg    Niels Bohr Institute

Copenhagen, Denmark

*ABSTRACT: During the Hipparcos mission in September 1992, I presented a concept for using direct imaging on CCDs in scanning mode in a new and very powerful astrometric satellite, Roemer. The Roemer concept with larger aperture telescopes for higher accuracy was developed by ESA and a mission was approved in 2000, expected to be a million times better than Hipparcos. The present name Gaia for the mission reminds of an interferometric option also studied in the period 1993-97, and the evolution of optics and detection in this period is the main subject of the present report. The transition from an interferometric GAIA to a large Roemer was made on 15 January 1998. It will be shown that without the interferometric GAIA option, ESA would hardly have selected astrometry for a Cornerstone study in 1997, and consequently we would not have had the Roemer/Gaia mission.*

### 1. Introduction

Only one astrometric satellite has been launched, Hipparcos, and its observations from 1989-93 brought a tiger leap of the accuracy and number of stars with good distances, proper motions and positions. In 2013, ESA will launch another large astrometric satellite Gaia, which is expected to bring a new tiger leap for astrometry. I have been deeply involved in both projects for 32 years from the very beginning of Hipparcos in 1975 when I made a completely new design of the satellite. I have written and lectured (Høg 2008, 2011) about these projects from my own perspective, but in a historically reliable manner with frequent checks of my memory by means of my archive and by correspondence with colleagues. I include personal recollections and reminiscences hoping to bring events and decisions closer.

In the summer of 1990 I began a collaboration with Russian colleagues about a successor for Hipparcos and we soon included Lennart Lindegren. This collaboration led to the Roemer proposal in 1992, to GAIA in 1993, and then to Gaia and gradually more and more people contributed to the development. *The chain of ideas and actions related to optics and detectors is my main subject*, not at all a complete history of Gaia.

---

[1] Contribution to the history of astrometry No. 11

In the spring of 2010, I realized that the role of the Roemer satellite proposal of September 1992 (Høg 1993) seemed to be forgotten. This proposal was important for two reasons, many of the new ideas in the proposal are contained in the final Gaia satellite and the proposal in fact started the work towards Gaia. By October 1993, only a year after presentation of the Roemer proposal with direct imaging on CCDs, the basis had been laid for the studies of Roemer and another option, GAIA, using Fizeau interferometers (Lindegren et al. 1993b).

Interferometric designs were studied in the years 1993-97, followed by a design without any interferometry, but based on the ideas in Roemer with direct imaging on CCDs from full-aperture telescopes. The mission thus became a Roemer mission with large telescopes. The name GAIA with the capital "I" for interferometry remained, however, until about 2003 when it was changed to Gaia. The CCD as a two-dimensional detector with high detection efficiency is better by many orders of magnitude than the photoelectric detector in Hipparcos which measured only one star at a time and this potential advantage of a CCD was trivial by 1990 when my design of a new astrometric mission began, the only question was how to do it with CCDs.

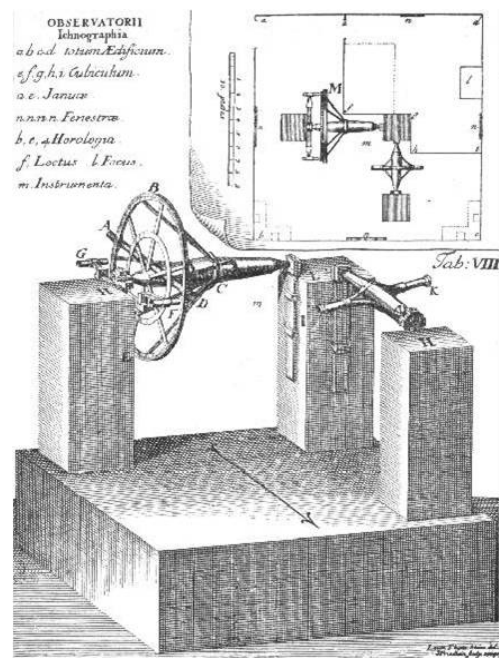

**Figure 1.** Ole Rømer (1644-1710) invented the meridian circle, the fundamental astrometric instrument for centuries.

I proposed the name "Roemer" as a proper name for an astrometric satellite. Ole Rømer (1644-1710) invented the meridian circle, see Fig. 1, the fundamental instrument of astrometry for centuries. He constructed many other physical and astronomical instruments in Denmark and he discovered the speed of light while observing in Paris.

Discussions and ensuing correspondence in 2010 led me to write about space astrometry plans in the 1990s, especially as related to Roemer and Gaia. I will begin with an account of what people are thinking or remembering of these things today, almost twenty years later. After an overview of Russian and American space astrometry follows the development of Roemer and Gaia in the 1990s. Then finally, very briefly, a view of the present Gaia design and of the future of space astrometry.

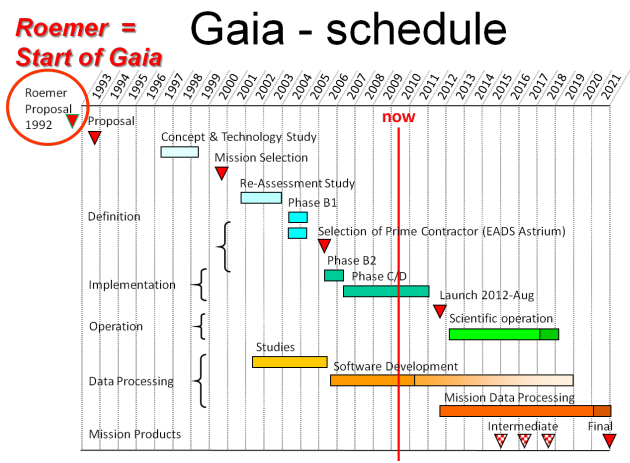

**Figure 2.** The Gaia schedule of mid 2010. I have inserted the time tag "Roemer proposal" at upper left.

## 2.  What People Thought in 2010

In the spring of 2010 I noticed that the time schedule for Gaia, Figure 2, began in late 1993. When I saw Timo Prusti, the Gaia Science Team leader, at the Gaia conference in Paris in June 2010 I suggested that he should recognise the role of Roemer by inserting a time tag at September 1992 with the words "Roemer proposal". This suggestion resulted immediately in remarks from several around the lunch table where we were seated that other suggestions had been made before 1992 saying that Hipparcos should be followed by another astrometric mission. I said that such suggestions are fine to make, but they did not trigger further work leading to Gaia. They were ideas or suggestions, not elaborate mission proposals as Roemer was.

Suggestions to use CCDs to measure hundreds of stars simultaneously had also been made it was continued at the lunch table. This is also an obvious idea, but not a mission proposal and this simple idea did not itself trigger the work that led to Gaia.

When I later mentioned this matter in a mail to several I received an offer from Francois Mignard. He proposed to include an image of Roemer on one of his slides if I wanted. This slide from his presentation in Paris as leader of the Gaia data reduction mentioned the "unfortunate followers" of Hipparcos aiming for 0.1 mas accuracy: "Roemer, FAME-1, FAME-2, DIVA, Lomonosov, AMEX". This was a kind offer by Francois, but not what I wanted. Here I want to stress that I do not blame Francois or anyone for not knowing so well what happened twenty years ago. How could I - when I see how much reading and correspondence was required to find out for myself. I want to thank Francois for many years of pleasant and efficient collaboration on astrometry and on the history of astrometry, including the present report.

Thus, in 2010 some participants in the Gaia preparations believed that merely general and vague ideas or suggestions had been made twenty years ago before the GAIA proposal and that the Roemer proposal was such a vague idea. This showed me that the history of Hipparcos-Roemer-Gaia and the relation to the many other mission proposals in the 1990s ought to be written. I contacted Michael Perryman, Lennart Lindegren and Ken Seidelmann and asked if they would collaborate in one way or another.

Michael answered immediately that he did not want to be involved, he felt no enthusiasm now about the history of Gaia. He continued: *"I did make my own extensive notes on the project, from its very beginnings, which cover the scientific process, the industrial design, the advisory committee politics, and many of the ESA internal issues. It runs (from memory) to some 40 pages of small text. ... Perhaps, in years to come, I will write my own recollections of the first 10 years. But not now."* Lennart and Ken kindly sent extensive comments in the ensuing correspondence which have been used in the following.

## 3.  American Space Astrometry 1990

American astronomers in the U.S. Naval Observatory (USNO) and elsewhere have before 1990 been engaged in astrometry from space especially with the Hubble Space Telescope (e.g. Duncombe et al. 1990 and Seidelmann 1990) which has in fact provided accurate milliarcsecond astrometry, especially after the optical repair mission in 1993.

American astronomers also had plans for high-accuracy astrometry by Michelson interferometers in space with pointed telescopes, POINTS, see Chandler & Reasenberg 1990 and Figure 3.

In both cases this was narrow-field astrometry, very different from the global wide-angle astrometry provided by Hipparcos, Roemer and Gaia. The Americans thought of pointed telescopes in a satellite for observation of a few thousand stars with very high accuracy while ESA astronomers were developing a scanning satellite for systematic global astrometry of a hundred thousand stars with milliarcsecond accuracy.

The difference between the American approach to space astrometry and that within ESA was one reason that no cooperation has resulted which left significant traces in the early 1990s in either ESA or USA in spite of much communication at conferences and otherwise. Another reason was that the astrometric expertise in Europe was so sufficient that no collaboration on Hipparcos had been needed. But the inspiration to use CCDs in scanning mode came to me from America, from the work in those years on a meridian telescope in Arizona described by Stone & Monet 1990.

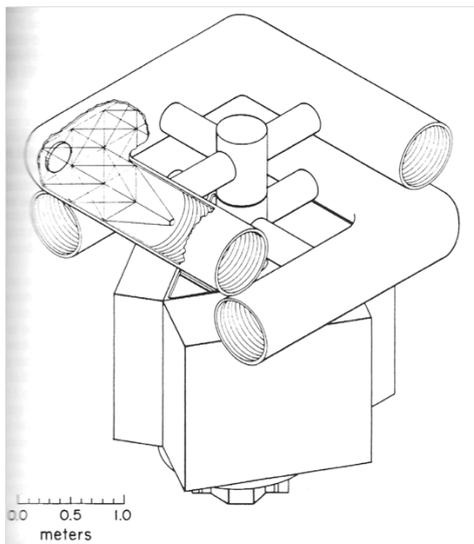

**Figure 3.** *"An artist's rendition of POINTS with 2-m separations between pairs of telescopes 25 cm diameter. The instrument, shown mounted on the Multimission Modular Spacecraft, comprises two U-shaped interferometers joined by a bearing that permits the angle between the principal axes of the interferometers to vary by up to a few degrees from its nominal value of 90 deg."* – quotation from original figure.

The idea from POINTS of Michelson interferometers in space for high-accuracy astrometry was further pursued in the 1990s and became the Space Interferometry Mission, also known as SIM Lite (formerly known as SIM PlanetQuest). It was a planned space telescope developed by the U.S. National Aeronautics and Space Administration (NASA), in conjunction with contractor Northrop Grumman. One of the main goals of the mission was the hunt for Earth-sized planets orbiting in the habitable zones of nearby stars other than the Sun. SIM was postponed several times and finally cancelled in 2010 (SIM 2011).

## 4. The Russian Collaboration

Inspiration to design a successor in space for Hipparcos could obviously not come to me from America. Nor could it come from the Hipparcos community where all attention was focused on the Hipparcos observations and data reduction. The inspiration came from Russia as I have described before (Høg 2007), to be partly repeated here with new details.

At a conference in Leningrad in 1989 we had heard about three plans in Russia, then USSR, for successors to Hipparcos. This became crucial for the development of Roemer and Gaia because I met an active interest in Leningrad and Moscow during the following years, especially with Mark Chubey and his team in the Pulkovo Observatory and with the Mission Control Centre in Moscow, without which there would have been no Roemer or Gaia mission today.

The three Russian plans were described at the IAU Symposium No. 141 held in Leningrad in October 1989: (1) Lomonossov with a pointing telescope of 1 m Ø, F=50 m aiming for 1 mas accuracy. (2) REGATTA-ASTRO: scanning telescope, 10 mas accuracy. (3) AIST shown in Figure 4: 2 telescopes, 0.25 m Ø, scanning, 1 mas.

All three aimed for launch before 1997 and Lomonossov and AIST expected 1 mas accuracy, similar to Hipparcos. The proposers considered as the primary scientific aim to get second epoch positions and thus very accurate proper motions for the 100 000 Hipparcos stars.

In fact, the inspiration to design a new mission came for me at a visit in the summer of 1990 to the Caucasus mountains with Mark Chubey and his team (see Figure 6), after I had lectured about Hipparcos and Tycho in Pulkovo and Moscow.

At first, during our discussions on the travel I just wanted to understand the Russian projects, especially the AIST, see Chubey, Makarov, Yershov et al. 1990 and Figure 4. But after a day's discussion I realized that I was thinking more about improving Hipparcos than about understanding how AIST was supposed to work.

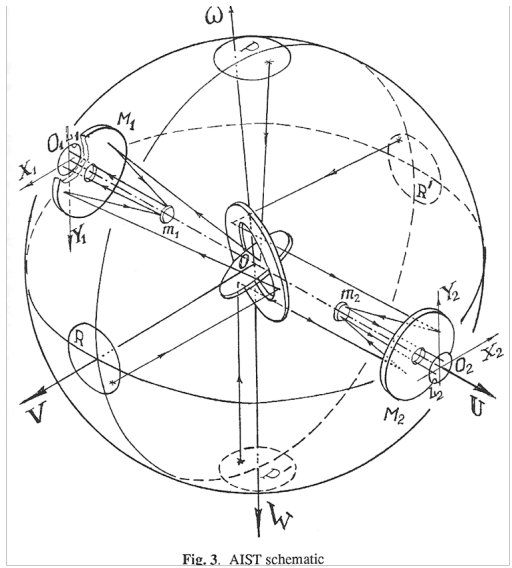

**Figure 4.** The proposed Russian astrometric satellite, AIST, with two telescopes.

Staying at the Kislovodsk Observatory in the Caucasus, I went out of bed in the first night and started to put my thoughts into a first drawing. I thought I was quiet, but Mark had heard me from his room and knocked my window asking in a concerned voice whether I was ill which I could truly deny. He then arranged for a better lamp at my table.

That was the morning of the first day with clear weather so that I could see the beautiful snow covered double peak of Mount Elbrus, an extinct volcano and the highest mountain in all of Europe. The day before they had sometimes pointed with the arm into the clouds saying: "Elbrus is over there". Now I understood why it was so important for them to point.

Our discussions were continued and one of the following meetings took place in Moscow in June 1991, now also with Lennart Lindegren present. It was important to involve Lennart in the design, he could make the correct estimate of the astrometric accuracy, always unfailing in mathematics.

I reported in the Hipparcos Science Team after every meeting with the Russians. Our leader Michael Perryman was a bit reluctant to give time, but he usually gave the 10 or 15 minutes I wanted, and useful discussion resulted. Michael was of course reluctant because it was his task as a leader to look after the observations and data reduction of the flying satellite Hipparcos, not to design a new mission, but he became very interested and active in such design in 1993.

At the Moscow meeting in June 1991 we presented ideas for a second Hipparcos, see Figure 5. The Hipparcos system, still with IDTs and photomultipliers, was improved with larger telescopes and enhanced star mappers. Expected accuracy was 1 mas for 400 000 stars and 0.2 mas/year proper motions for the 120 000 Hipparcos stars. We wrote: "The proposal … is being considered by the Mission Control Centre, Moscow."

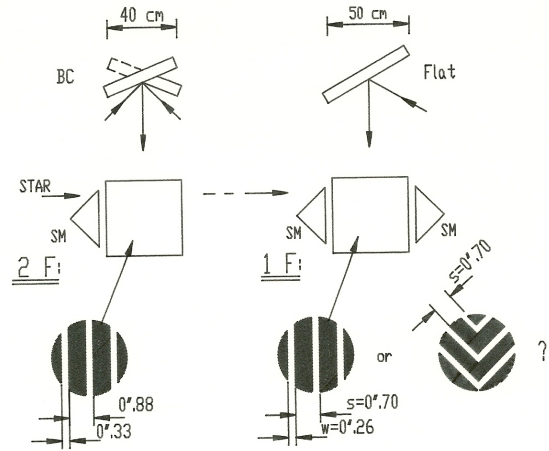

**Figure 5.** A second Hipparcos with two telescopes, proposed in 1991 by Høg & Chubey. The 2F telescope observes two fields by means of a beam combiner. The 1F telescope with larger aperture has only one field of view. The detectors are still photoelectric as in Hipparcos.

The paper was accepted for publication, but the proceedings never appeared. It was also presented as a poster at the IAU General Assembly in Buenos Aires in August 1991, but was not accepted for publication. It is now scanned and placed on my homepage, Høg & Chubey 1991.

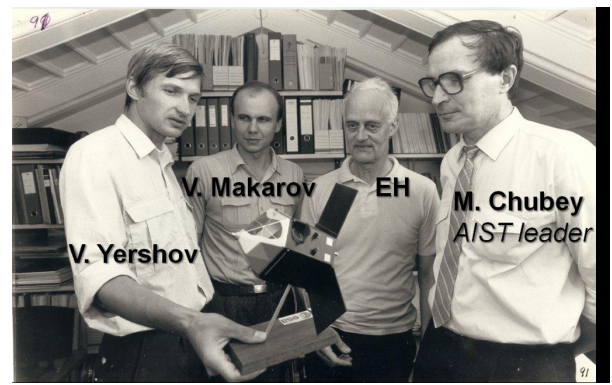

**Figure 6.** The Russian space astrometry team visited Copenhagen in 1991 and is here looking at the Hipparcos model. One member, Valeri Makarov, then stayed seven years in Copenhagen working on the Tycho-1 and Tycho-2 Catalogues.

After this exercise my study began of how to use a CCD (Charge Coupled Device). This two-dimensional detector was invented in 1969 by W. Boyle and G.E. Smith who subsequently received the Nobel Prize for physics in 2009. In 1979 an RCA 320x512 pixel cooled

CCD system was first used on a 1-meter telescope at Kitt Peak National Observatory. This started the take-over from photographic plates in astronomy, including astrometry, and a similar take-over from films in hand-held digital cameras began ten years later.

Two proposals were made for the use of 2D detectors in Hipparcos according to mails in 2011 from M. Perryman and R. Le Poole. A proposal to use a solid-state two-dimensional detector, an ICID – not a CCD, in Hipparcos was made by di Serego Alighieri et al. 1980. Another proposal by M. Hammerschlag in 1981 transported the charges in a CCD perpendicular to the motion of the stars, while Gaia transports along with the motion of the stars in TDI mode (Time Delayed Integration). By 1990 when my work towards Roemer and Gaia began the potential advantage of using a 2D solid-state detector was trivial, the only question was how to do it. The two proposals were in fact not known to me at that time.

The higher quantum efficiency of a CCD and the ability to observe many stars simultaneously would be the great and very obvious advantage over the photoelectric detectors. I learnt in 1991 from our electronics engineer in Copenhagen, Ralph Florentin Nielsen, what a CCD can do and what it cannot do.

But there was doubt in those years about the use of CCDs for astrometry. Their dimensional stability was doubted, the sensitivity was perhaps not uniform over the individual pixels, and the position of the pixels was perhaps not stable and had to be calibrated.

With these concerns in mind I began with a design using the CCD as a modulation detector. We knew from Hipparcos that a very accurate grid could be manufactured and would be very stable, so that seemed to be the way to go and out came the design in Figure 7. We could see that the astrometric efficiency of the satellite would be 1000 times higher than that of Hipparcos, as always based on Lennart's calculations.

## 5. Roemer and GAIA 1992-1994

The proposal with a CCD as modulation detector was called Hipparcos-2 in a report of January 1992. It was submitted to the IAU Symposium No. 156 which was to be held in September 1992 in Shanghai. But in May my study began of direct imaging on CCDs and this soon promised to be a hundred times better than the system with modulation so I thought it deserved the name Roemer, not just Hipparcos-2. We brought both manuscripts to Shanghai and both were accepted for oral presentation (Høg & Lindegren 1993, and Høg 1993, respectively.)

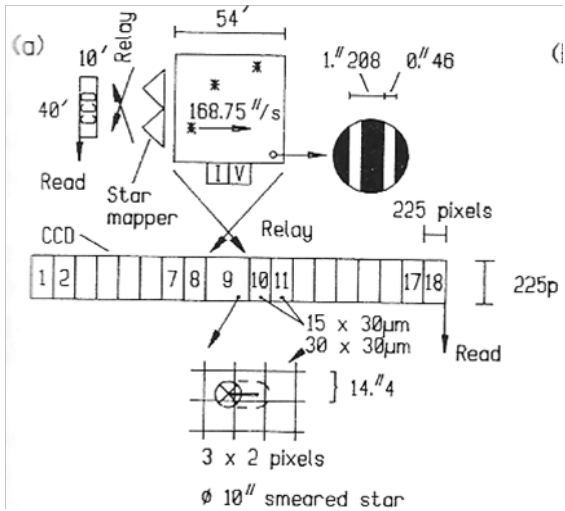
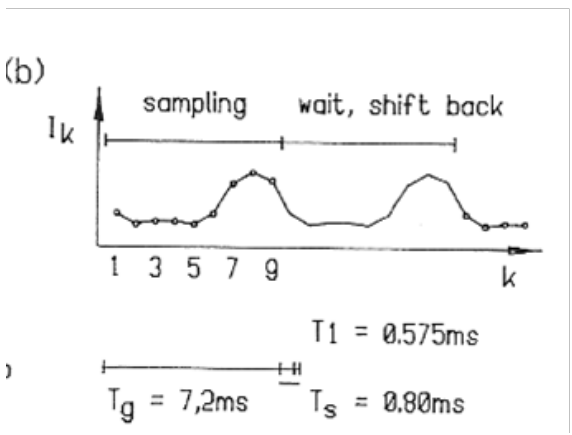

**Figure 7.** The detection of modulation with a CCD was proposed in 1992 for a Hipparcos-2 satellite (Høg & Lindegren 1993).

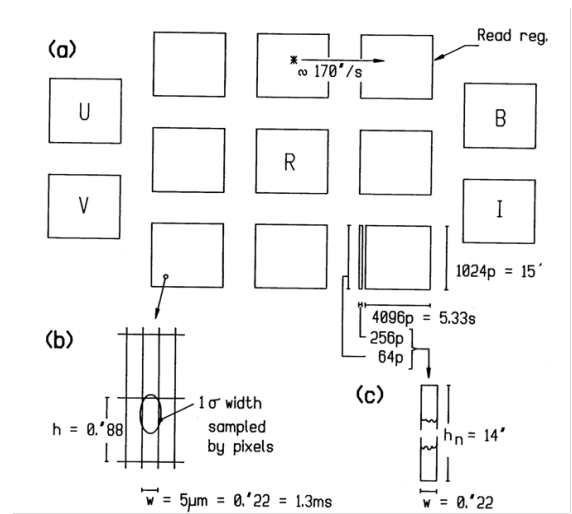

**Figure 8.** Focal plane of the Roemer satellite with CCDs in scanning mode proposed in 1992, stars moving left to right through the field (Høg 1993).

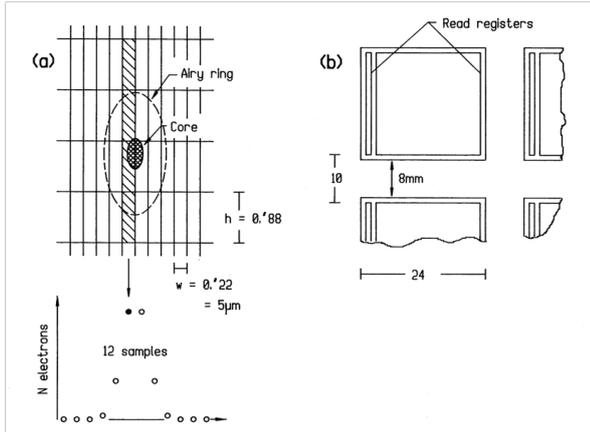

**Figure 9.** Sampling of the Roemer CCDs. (a): The pixels are elongated perpendicular to the scanning, several pixels (here 4) are read together in order to decrease the readnoise, and (b): short CCDs are provided for bright stars, features later adopted in Gaia.

Roemer was a very specific mission proposal with CCDs in time delayed integration and with direct imaging of the stars on the CCDs (Høg 1993, see Figures 8 and 9). It was a scanning satellite with a beam combiner similar to that in Hipparcos. For a 5 year mission an astrometric accuracy of 0.1 mas was predicted at V=12 mag, more than 10 times better than Hipparcos. The astrometric efficiency was 100 000 times higher, but obtained with the same telescopic aperture of 29 cm as Hipparcos. Astrometry and multicolour photometry for 400 million stars were included.

The use of a CCD directly in the focal plane is astrometrically much more efficient than a modulating grid as in Hipparcos because much light is lost in the grid. The use of many CCDs with their higher quantum efficiency than the photoelectric IDT detector of Hipparcos, smaller transmission losses than in the IDT relay system, and the capability to observe thousands of stars simultaneously translates into at least 100 000 times higher astrometric efficiency for the same telescope aperture.

Further improvement of accuracy through larger aperture would come from the higher angular resolution and from the larger number of photons collected. This path was chosen for the Gaia mission development after the studies from 1993-1997 had shown that interferometry was not the way to go, as shall be elaborated below.

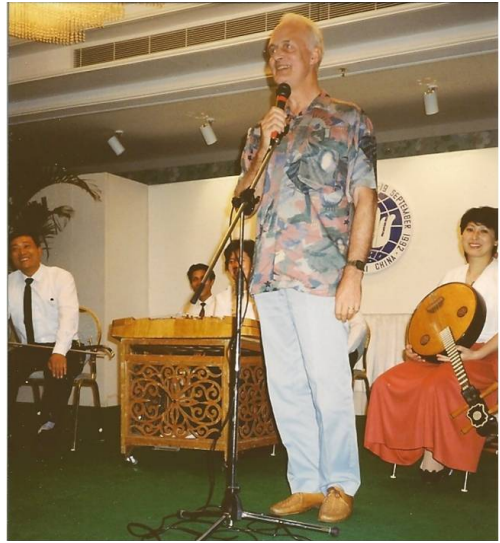

**Figure 10.** Shanghai 1992. You are encouraged to entertain at the conference dinner, I did so in Chinese and other languages.

During the conference in Shanghai three of us, Jean Kovalevsky, Ken Seidelmann and myself, agreed to apply for an IAU Symposium dedicated to sub-milliarcsecond optical astrometry. The symposium was approved and held in The Hague, August 15-19, 1994 (Høg & Seidelmann (eds.) 1995) and we wrote in the preface to the proceedings: *"Astrometry is on the threshold of great changes due to the fact that this decade, alone, is witnessing an improvement of stellar positions equivalent to the total improvement of the previous two centuries."*

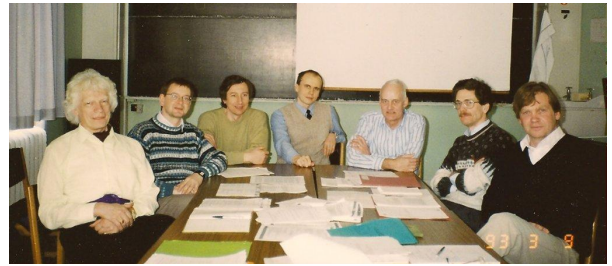

**Figure 11.** Roemer proposers met in March 1993 in Copenhagen: Kovalevsky, Lindegren, Halbwachs, Makarov, Høg, van Leeuwen, Knude – missing here: Bastian, Gilmore, Labeyrie, Pel, Schrijver, Stabell, Thejll.

The ideas in Roemer were adopted in a mission proposal submitted to ESA on 24$^{th}$ May 1993 for the Third Medium Size ESA Mission (M3). We proposed to measure 100 million stars and to obtain an accuracy of 0.2 mas at V=13 in a 2.5 year mission with a 34 cm telescopic aperture. Some of the proposers were members of the Hipparcos Science Team (see Figure 11 and Lindegren et al. 1993a).

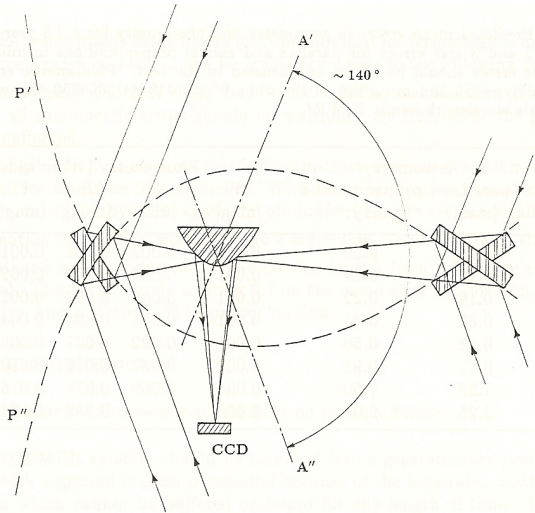

**Figure 12.** *"Optical sketch of a Fizeau-type scanning interferometer. The beam combiners are part of confocal paraboloids (P', P'') whose axes A' and A'' make a fixed angle of approx 140 deg."* – quotation from the original figure of 1993.

The mission was in fact called ROEMER, an acronym for "Rotating Optical Observatory for Extreme Measuring Efficiency and Rigour", but we will here simply call it Roemer. A section called "The FIZEAU option", not part of the baseline proposal, was included *"to point out a possible improvement towards a scanning satellite with ten times the angular accuracy of ROEMER"*. The section described a scanning satellite with *"two confocal Fizeau-type (or 'wide field') interferometers whose axes form a basic angle of the order 140 deg."*, a description fitting very well to the later GAIA. But the included optical system, shown here in Figure 12, underwent major development before it was called GAIA, an acronym for Global Astrometric Interferometer for Astrophysics.

The Roemer proposal was also presented at a conference in Cambridge in June 1993 (Høg and Lindegren 1994).

The proposal to ESA was rated by the Astronomy Working Group (AWG) to be the best among all astronomical proposals for M3. But it was considered to come too soon after Hipparcos and it was not sufficiently ambitious with respect to accuracy. It was therefore referred to a Cornerstone Mission study if 10-20 μarsecond accuracy could be demonstrated.

The proposal of a Cornerstone study meant that the AWG members got rid of a competitor for the M3 mission, but we should be grateful that we did not get approval for M3 since that would have prevented us ever to design the much more powerful Gaia.

As a reply to an ESA call for proposals of Cornerstone studies we submitted on 12 October 1993 a proposal to study for astrometry "a large Roemer option and an interferometric option", GAIA. They should be studied as two concepts for an ESA Cornerstone Mission for astrometry "without a priori excluding either", as Lindegren wrote in the cover letter.

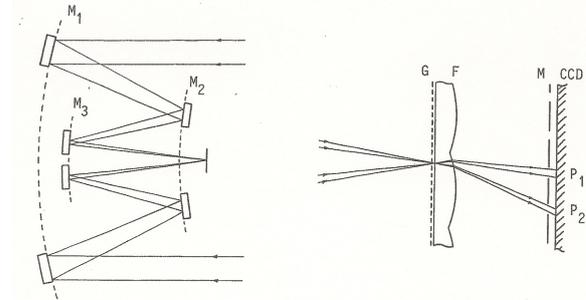

**Figure 13.** The GAIA system as it appeared in October 1993 (Lindegren et al. 1993b). Fizeau interferometer at left, modulating grid (G) with field lenslets (F) and detector (CCD).

The interferometric concept based on the above FIZEAU option was included in the report by Lindegren et al. 1993b, a concept which was mainly a result of discussions between Lindegren and Perryman. Two features in Figure 13, compared with Figure 12 should be mentioned. Firstly, the parabolic mirrors, which would give only a very small usable field, have been replaced with a three-mirror telescope, and secondly, principles of a CCD detection system are indicated.

In September 2010 Lindegren wrote to me: *"... in April 1994 during the HST meeting in Lund (14-15 April), I did write (in consultation with Michael) an e-mail to Steven Beckwith, the chairman of the UV-to-radio topical team of the Survey Committee then drafting ESA's Horizon 2000+ plan. In the e-mail, which was copied to L. Woltjer (chairman of the Survey Committee),* **I again stressed that Roemer and GAIA should not be seen as competing projects** *but as an indication of the different ideas circulating in the community, and the strong conviction that an advanced astrometric mission would be technically feasible and extremely worthwhile",* quoted from Lindegren 1994.

Development of the ideas was continued, especially by Lennart Lindegren, Michael Perryman and myself, of the proposed two mission concepts with higher accuracies: First, the interferometric mission GAIA, (see Figure 14) and second, a Roemer mission, called Roemer+, (see Figure 15) with larger apertures by Høg in August 1994 (Høg 1995).

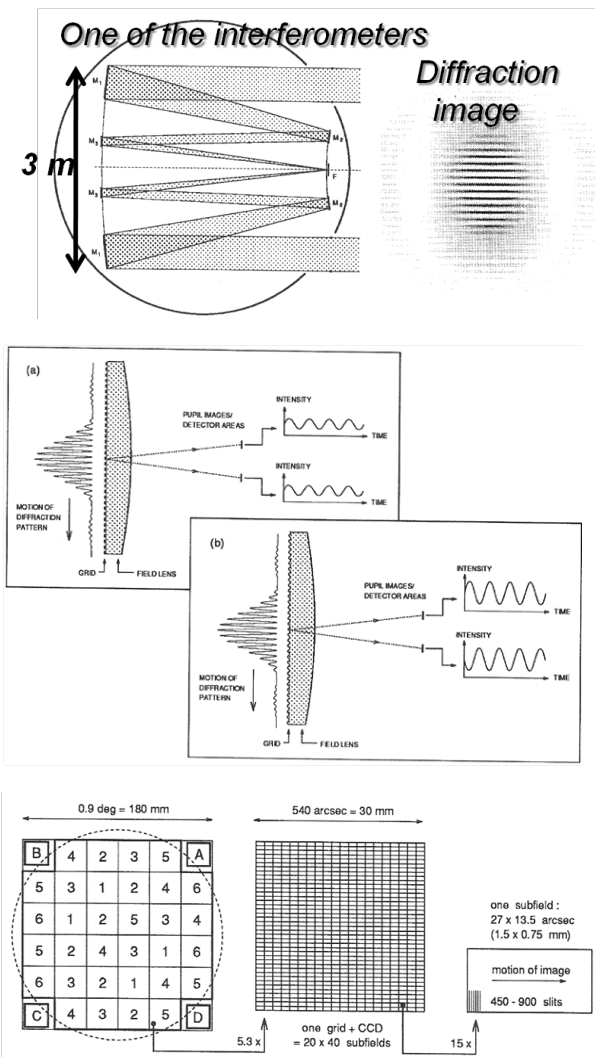

**Figure 14.** The GAIA system as it appeared in August 1994 (Lindegren & Perryman 1995). The optical system at top, then detection of the modulation, and more details below.

Thus, the basis for the studies of Roemer and GAIA was laid within one year and expanded with the large Roemer+ in August 1994 which could reach the 10-20 μarcsec goal. But by 2010 the role of Roemer had been forgotten, although the present Gaia may be seen as a large Roemer.

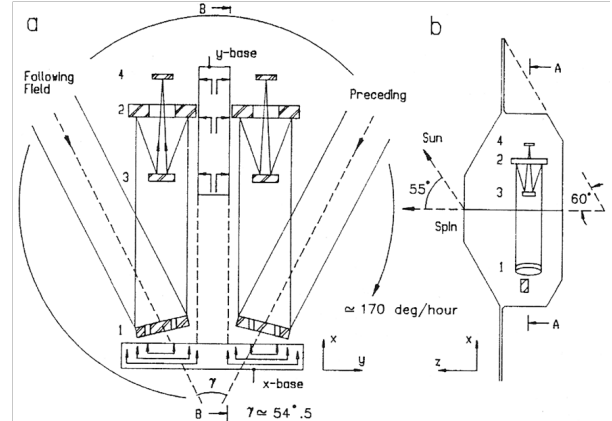

*Figure 2.* The Roemer+ satellite design. (a) Two Baker-Schmidt telescopes with tilted reflective corrector plates are pointed at the scanning great circle. The optical components are monitored by interferometric distance gauges with picometer precision, thus obtaining the variations of the basic angle as function of time. (b) Section of the rotationally symmetric satellite.

**Figure 15.** The Roemer+ satellite design of August 1994 (Høg 1995a), the first large Roemer. The use of picometer sensors is indicated.

## 6.  Studies of interferometry 1993-1997

Ken Seidelmann writes to me on March 10, 1994 that USNO is pursuing NEWCOMB which is however unfunded, and that they are interested in collaboration on other space-based astrometry projects like Roemer and GAIA.

In August 1994 American astronomers presented the Newcomb Astrometric Satellite, *"a concept for a small, quick, inexpensive, initial optical interferometer in space"*. It *"would have a stacked set of 3, or 4, Michelson optical interferometers..."*. It would be a pointing satellite with a precision of 0.1 milliarcsecond. Requirements, but no specific design was included." Quotations are from Johnston et al. 1995.

In early 1995 they began to think about wide angle astrometry with a scanning satellite. They proposed FAME with Fizeau interferometers as a MIDEX mission of NASA (Johnston 1995a and 1995b), but did not succeed.  This proposal is called FAME-1 or the "first FAME design" in the following to distinguish from the FAME-2 proposal a few years later.

In July 2010 Ken Seidelmann wrote: *"Ken Johnston and I started with the Newcomb proposal which evolved into the first FAME design, which was a Fizeau interferometer with a JPL optical design. That proposal was submitted to NASA. I gave a presentation in Europe on that design…"* It was in Cambridge in June 1995 (Seidelmann et al. 1995).

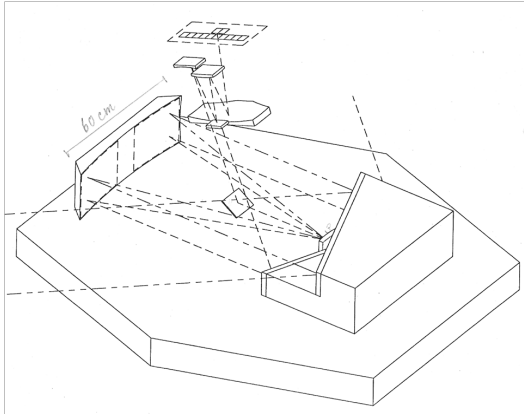

**Figure 16.** The FAME-1 optical design with Fizeau interferometer of September 1995.

I was a member of the FAME team from May to about November 1995 on invitation by Ken Johnston, but my duties did not allow me to continue. In October 1995 Johnston proposed to visit Copenhagen to meet with Lennart, Floor, and me to discuss the FAME proposal, but we answered in November that we were too pressed for time since Hipparcos and Tycho were still occupying us. Likewise we were too busy to offer a collaboration on the data reduction.

In September I received an improved optical system shown in Figure 16, also included in Johnston 1995a. That led me to design a new system, GAIA95, in October 1995, also a Fizeau interferometer but built into a Gregorian telescope, see Figure 17. A prism placed at the intermediate exit pupil lets the light through a hole in its middle. The focused but almost parallel beam returned from the secondary S3 passes through the prism thus providing spectra of all stars perpendicular to the scan direction. Astrometry and photometry could be obtained from the same images, the dispersed fringes. The system provided imaging without any disturbing central obscuration, utilizing the space between the two beams of the Fizeau interferometer.

The system was studied in Copenhagen and a report was distributed soon after by Høg, Fabricius & Makarov 1995. The publication by the same authors appeared in 1997.

The GAIA95 system with a D=1.5 m primary is shown in Figure 17. It could provide 10 µarcsec precision at V=14 mag and was considered for GAIA in 1997.

Soon after the first report had been distributed in 1995 I was called by Uli Bastian from Heidelberg. He asked if I could keep a secret for a few months. I promised and he continued saying that I could look forward to a small satellite on the GAIA95 design being launched before my 70$^{th}$ birthday - which would be 17 June 2002.

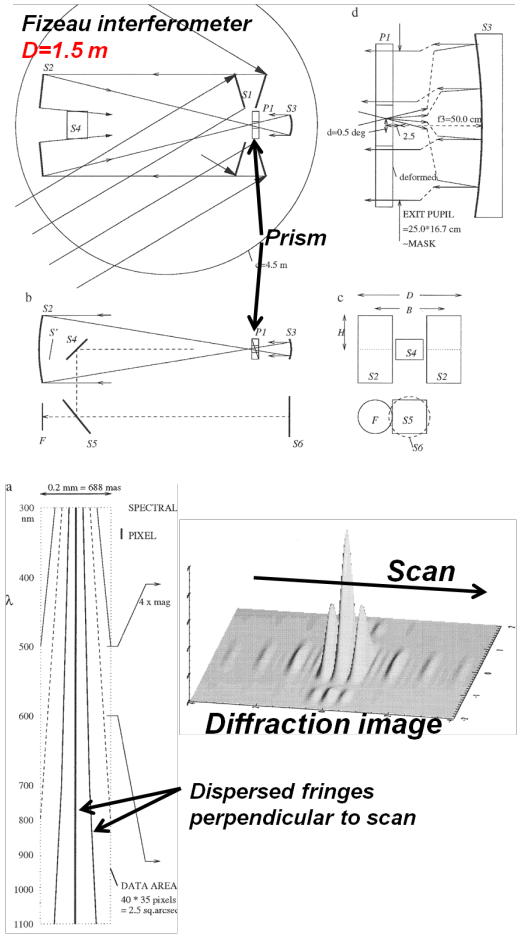

**Figure 17.** The GAIA95 design of October 1995 by Høg, Fabricius & Makarov, a Fizeau interferometer with Gregorian telescope, providing dispersed fringes for simultaneous astrometry and spectrophotometry.

Uli and German colleagues were working on a project which became the German interferometric minisatellite for astrometry and photometry, DIVA, Röser et al. 1996. The DIVA telescope was scaled to one tenth of GAIA95, i.e. to an aperture D=15 cm. In the fall of 2000 it was officially selected as a minisatellite project by the German space agency DLR. It was abandoned in 2002 when one of the German funding partners had dropped out. From 2002 to 2003 a follow-up collaboration with the USNO was tried, partly under the name of AMEX. This effort ended in 2004 when NASA funding did not materialise, and German astronomers decided to focus on the GAIA mission.

Ken Seidelmann wrote recently that many of the considerations concerning Gaia were going on at roughly the same time for the scanning satellite FAME2 (Johnston 2003), e.g. was the interferometer dropped in 1997 or 98. It was approved as a NASA MIDEX project in 1999, but was cancelled in 2002 primarily due to CCD chip production problems and budget concerns.

But these developments do not belong to our subject. The present report is about the development of the various optics and detector systems of Roemer, GAIA, and Gaia, including the scientific and technical environment for a better understanding. It is not a history of the Gaia project, nor is it of course a history of space astrometry in total. An overview of "future astrometric missions" is given in August 1996 by Seidelmann (1998). Seidelmann briefly describes ten plans (one by ESA, one in Germany, one in Japan, three in Russia, and four in the USA) of which one has survived, the interferometric GAIA, *"with a possible launch date of 2015."*

I was a team member of DIVA at the beginning, but then concentrated on the GAIA development. Michael Perryman insisted, and rightly so, that a member of the GAIA team could not also be member of another space astrometry team. The GAIA team had the task of developing a project for ESA, and should not be a centre for development of astrometric satellites.

The intense work on space astrometry development during these years appears from the listed proceedings of four international astrometric meetings: 1993 in Cambridge, UK, 1994 in The Hague, 1995 again in Cambridge, and 1997 in Venice. References are given below to the 22 papers especially on the Roemer and GAIA optics and detection systems. We find only one paper of this kind in 1993, two in 1994, 13 in 1995 and six in 1997. The GAIA95 option is only one example from the development of optics and detection for the interferometric GAIA in those years, but rather interesting because of the connection to American space astrometry and to DIVA.

Among the 22 papers, only four discuss the non-interferometric full-aperture Roemer option (Høg & Lindegren 1994, Høg 1995a, and 1995d, and Yershov 1995.) We scientists did not follow Lindegren's recommendation to ESA in 1993 and 1994 to study the Roemer and GAIA options "without a priori excluding either".

We were all very fascinated by the idea of Fizeau interferometry and we worked hard on its development though still occupied primarily with the data reduction and publication of Hipparcos and Tycho results. But our opinion changed when the ESA Cornerstone study started and the ESA Science Advisory Group, led by Michael Perryman, could discuss the studies by the industrial and ESA teams. We saw then that interferometry was the wrong track and we returned to direct imaging on CCDs in telescopes with apertures as large as could possibly be contained in the ESA launcher.

### 7. Cornerstone study approved 1997

The capital letter "I" in GAIA stood for Interferometry since the proposal in 1993 and the name was maintained although interferometry was dropped already in January 1998, but about 2003 the name was changed from GAIA to Gaia. In 2007 I proposed to change the name from Gaia to Roemer (Høg 2007) in recognition of Ole Rømer, this proposal of course came too late.

The names GAIA or Gaia have been maintained throughout the years for the sake of continuity and because we did not want to draw attention to the fact that interferometry had been dropped since we knew that ESA had attached importance to the use of interferometry when the Cornerstone Mission study was approved.

This appears from the letter of invitation to join the Science Advisory Group (SAG) for an Astrometry Cornerstone. The letter of invitation (ESA 1997) is dated 11 March 1997 and reads:

*"Space interferometry was identified in the ESA long-term programme for space science, Horizon 2000, as a potential candidate among space projects planned for after the turn of the century.*

*Recently, a Survey Committee established by ESA, has updated this programme as the Horizon 2000 Plus plan, which identifies three major projects over the period 2006-2016. One of these is an interferometry observatory as a Cornerstone mission open to the wide scientific Community. The first option would be to perform astrometric observations at the 10 µarcsec level. As an alternative option, the Survey Committee recommends studies of infrared interferometry, in particular with the aim of detecting planets around other stars. ... in January 1997, ESA's Space Science Advisory Committee recommended to start the study activities in preparation for the future definition of these interferometry projects. ..."*

The Cornerstone study of infrared interferometry mentioned led to the project Darwin. On the ESA Portal I now looked for "interferometer" and found a page from August 1997 about Cornerstones for GAIA, Darwin, and LISA. Asking for Darwin the answer was a page beginning with *"Study ended 2007, no further activities planned"*.

It appears that astrometry by interferometry is mentioned in the invitation, but neither GAIA nor Roemer. It seems therefore, that astrometry would not have been chosen by the Survey Committee for a Cornerstone study if our proposal had only contained a large Roemer and if no interferometry had been included. We were cautious in the Science Group never to make a point of the fact that interferometry was dropped after less than a year of study by industry. We continued with the Roemer option reasoning that ESA wanted the astrometric science and not a particular instrument.

But today we should lay open what really happened. (1) Without the Roemer proposal of 1992 there would have been no GAIA proposal in 1993, (2) without the GAIA with interferometry no selection for a Cornerstone study in 1997, and (3) without the Roemer concept of direct imaging with full-aperture telescopes there would not have been a feasible astrometry mission to approve in 2000 since the approved GAIA or Gaia may be seen as a large Roemer mission with many of the features proposed for Roemer in 1992.

It would be interesting to know what happened in the Survey Committee and the SSAC in order to understand why they selected interferometry. The minutes of the meeting must still exist and some participants could be interviewed. It seems clear that they were not asking for the best possible astrometry because they ignored the clear recommendation of the astrometry experts in the proposal by Lindegren et al. 1993b and repeated in the letter by Lindegren 1994. The recommendation was to study *"a large Roemer option and an interferometric option ... without a priori excluding either."* The ESA committees had a "great vision of interferometry" rather than a vision of great astrometry.

In an ESA committee of astronomers most members will be astrophysicists. They will often consider astrometry to be very useful for astronomy, but when it comes to a decision between expensive projects, astrometry has a very difficult standing. This was the case at the approval of Hipparcos as I have shown in Høg 2011a: *"Miraculous approval of Hipparcos in 1980".* Several miracles happened then. Miracles only happen when good and strong persons take action.

## 8. From GAIA to Roemer/Gaia

The approval of a 'Concept and Technology Study' for GAIA (along with the other three cornerstone mission candidates) was given in 1996 and an ad hoc Science Advisory Group (SAG) was established in March 1997. *"A one-year industrial study took place between mid-1997 and mid-1998. Three industrial proposals were submitted in June 1997. The contract subsequently was awarded to Matra-Marconi Space (MMS) in July 1997"*, quotation from ESA 2000.

The GAIA SAG had its first meeting in March 1997 led by M. Perryman and the members were: F. Mignard, P.T. de Zeeuw, G. Gilmore, E. Høg, M. Lattanzi, L. Lindegren, and S. Röser – K.S. de Boer and X. Luri joined the SAG later. The following three years of work resulted in the Concept and Technology Study ESA 2000 which presents the scientific case of GAIA on 100 pages and the technical design, mission performance, and data analysis on a further 300 pages.

From these years of intense work I shall here only mention some of the main steps in the design of the payload. I will describe only few individual contributions, but I want to emphasize that Michael Perryman was our very efficient and competent leader all the time – without Michael Perryman and Lennart Lindegren there would be no Gaia.

The first design, Figure 18, from June 1997 corresponds to the GAIA proposal by Lindegren et al. 1993b with stacked Fizeau interferometers. A separate telescope, ARVI, for radial velocities has been included as proposed by Favata & Perryman 1995.

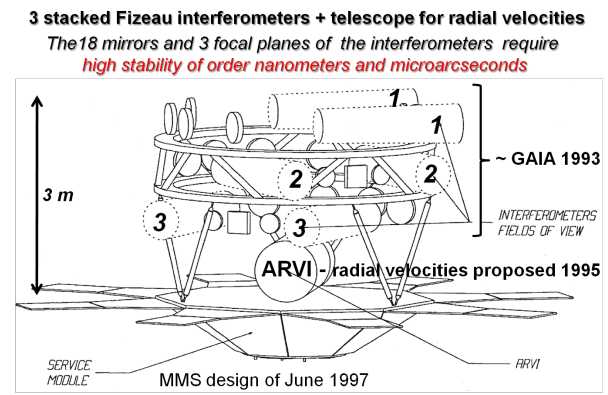

**Figure 18.** The GAIA design of June 1997 as it appeared from the study proposal by MMS (Matra Marconi Space). The detection system with or without a modulating grid, shown in Figures 13 and 14, was dropped by MMS a few months later and by ESA in January 1998 and only detection of the diffraction image directly on the CCD with full telescope aperture was further considered.

It would be interesting to see when the decisions about various important features in the instrument were taken, but it is usually not possible to give a precise time because of the many interrelated issues considered simultaneously. For example, the pros and cons of having three stacked interferometers, as in Figure 18, or only two were discussed at the 4[th] SAG meeting held in Grasse on 24 September 1997, based on a report by Lindegren. Two interferometers remained the baseline.

At the same meeting the requirement of the 90 cm aperture of ARVI was confirmed, with a possible target of 1.3m. MMS was investigating inclusion of ARVI within the interferometer assembly.

I continue to read from the careful minutes by Perryman of this meeting: Technically, the mission target was now a billion objects to a completeness limit of 20 mag. Further studies of the possible photometric system had been made by Høg, with plans for further independent activities to be carried out by Munari (Padova).

Høg presented his latest concept of the sky mapper, sending down information on specific objects. Gilmore presented the case for sending down the entire sky data, which would probably be feasible now with the present sky mapper approach and realistic compression schemes.

Work on on-board detection instead of using an a priori GAIA input catalogue, similar to the Hipparcos input catalogue had already begun in July 1997.

Lattanzi presented the case for a beam combiner (instead of separate interferometers) being studied by Alenia/Aerospatiale/APLT/AMTS.

An outline of the final report, ESA 2000, was presented by Michael Perryman. This is typical for Michael, he was always very timely with preparations for everything, be it international meetings, the team meetings, the subsequent minutes, and in this case with the outline for the final report about the work we had just begun. The outline was in front of us already two years before the report should be completed and this was of course very important for a report which finally contained 381 pages, ESA 2000, the report so crucial for obtaining the mission approval in 2000.

Michael evidently worked all the time on our GAIA project as he had done on Hipparcos since he became leader of that project in 1980. I could always count on him, e.g., to have a long phone conversation after a SAG meeting in order to discuss an issue which had not been adequately solved during the meeting. He could very quickly grasp the essence of any scientific or technical problem. He always arranged a social dinner after the first day of our two-day meetings including engineers from ESA and the industry and I thoroughly enjoyed these dinners with talk of science and many other things. His arrangement of our visit to the Hipparcos launch in South America in 1989 is unforgettable.

The industrial mid-term review was held in ESTEC on 14 January 1998, followed by the 6$^{th}$ SAG meeting of two days. Frédéric Safa, the leading MMS engineer, *"summarised the main lines that the instrumental development had followed over the last two months: (a) the primary was now baselined as a 1.7 x 0.7 $m^2$ monolithic reflector, with an overall f=50m, and resulting in a pixel size along scan of 9µm. (b) the passive telescope design could be achieved without the requirement of nm-accuracy mechanisms. MMS/Safa presented their concepts for the measurement (not control) of basic angle variations at or below the 1 µarcsec level. ... "*

The minutes of the meeting shown in Figure 19, gives the many strong arguments for the monolithic full-aperture reflector, i.e. for abandoning interferometry. Industry, not SAG scientists, had now studied "a large Roemer option", and industry would certainly have found this solution even if we had not proposed the large Roemer to ESA in October 1993. The SAG agreed to the new baseline without anybody thinking of Roemer as far as I know, not even I thought of Roemer. The transition from GAIA with interferometers to a large Roemer can be fixed in time precisely to 15 January 1998.

The last lines in the text of Figure 19 were followed with the inclusion of an interferometric option of Alenia design in the "Red Report", i.e. ESA 2000 - which in fact has a white cover.

---

Extract from the minutes of the SAG meeting on 15 January 1998: *Filling of the central aperture allows the light collecting power to be preserved while decreasing the overall diameter of the payload and spacecraft. Utilization of the 9µm rather than the 6µm pixels allows the payload feasibility to be considerably eased. The monolithic primary allows the structural properties to be enhanced, and leads to a lower data rate (and improved readout noise performance etc) being achieved. The major 'disadvantage' of this approach is that the mission becomes less obviously interferometric. However the central issue is that given the performance figures now achieved with the compact, monolithic, 9µm system, enhanced performance by insisting on an interferometric payload would only come with penalties of detector technology, mass, dimensions, structural stability, and cost, with limited accuracy improvements. At present it is difficult to envisage how such an 'artificial increase' in complexity could be justified. For the final Red Report, these issues could be addressed, with an interferometric option being included for illustration. In parallel, the Alenia design which might be included as an alternative option in the Red Report, is more evidently following more classical interferometric principles.*

**Figure 19.** Arguments for the new GAIA design of January 1998 without interferometric telescopes.

Today, the following reasoning is natural. It is rather trivial that a full-aperture large mirror gives more information of the image position than if the middle part of the large mirror is removed in order to make a Fizeau interferometer. This was put in mathematical form by Lindegren (1998). The author shows: *"... why a single optical aperture is better for a scanning satellite than the originally proposed twin pupil of a Fizeau interferometer."* You cannot improve on astrometry by removing some photons. Secondly, the great complication and consequent high cost of a large interferometer is also obvious considering the requirements on stability of relative positions and angles of many mirrors on the order of nanometers and microarcseconds, cf. Figures 13 and 18. Thirdly, the three stacked interferometers would require an enormous shield against the Sun.

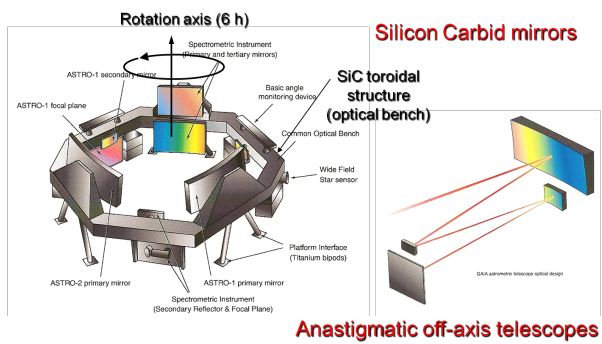

**Figure 20.** The GAIA design of mid 1998. Interferometry has been dropped and the design has become a large Roemer.

These three handicaps make the interferometric GAIA a certain loser against the large Roemer - so it appears with hindsight. But this was not at all clear to me in those years and we never discussed these issues. I worked on the interferometric option, e.g. on optics with GAIA95 and on the detection, Høg 1995c. We all worked with enthusiasm because we believed interferometry could give good astrometry, not just for tactical reasons knowing that interferometry was fashionable in ESA and elsewhere. All of us believed in an interferometric GAIA. We left to the industry to solve the problems we had vaguely seen and the engineers quickly opened our eyes in January 1998.

The design by mid 1998 at the end of the industrial study is shown here in Figure 20. A contemporary status of the GAIA project was presented at the meeting in Gotha in May 1998, available as Lindegren 1998 and Høg et al. 1998. The satellite contains two large telescopes for astrometry instead of three interferometers and a separate telescope of 0.75 x 0.70 m² aperture for radial velocities and photometry. Photometry in four broad bands is obtained in the astrometric telescopes and in seven medium-width spectral bands in the smaller telescope.

The design is described in great detail in ESA 2000 where the number of intermediate bands has been increased from seven to eleven. The GAIA mission was approved by ESA in 2000 for a launch "not later than 2012." A GAIA Science Team was set up, again led by Michael Perryman. The members up to 2007 were:

| | |
|---|---|
| Frédéric Arenou *(2001 - 2005)* | Meudon, France |
| Carine Babusiaux *(2006 - 2007)* | Meudon, France |
| Coryn Bailer-Jones *(2001 - 2007)* | Heidelberg, Germany |
| Ulrich Bastian *(2001 - 2007)* | Heidelberg, Germany |
| Anthony Brown *(2006 - 2007)* | Leiden, The Netherlands |
| Mark Cropper *(2006 - 2007)* | MSSL - UCL, United Kingdom |
| Erik Høg *(2001 - 2007)* | Copenhagen, Denmark |
| Andrew Holland *(2001 - 2005)* | Leicester/Brunel, United Kingdom |
| Carme Jordi *(2002-2007)* | Barcelona, Spain |
| David Katz *(2001 - 2007)* | Meudon, France |
| Mario Lattanzi *(2001 - 2005)* | Torino, Italy |
| Floor van Leeuwen *(2003 - 2007)* | Cambridge, United Kingdom |
| Lennart Lindegren *(2001-2007)* | Lund, Sweden |
| Xavier Luri *(2001 - 2007)* | Barcelona, Spain |
| Francois Mignard *(2001-2007)* | Nice, France |
| Fred Jansen *(2006)* | ESA/ESTEC (Gaia Project Scientist) |
| Michael Perryman *(2000 - 2006)* | ESA/ESTEC (Gaia Project Scientist) |

All the years were busy for my own part with work on many aspects of GAIA, for instance on the design of the optimal photometric system, after 2000 in the chair of the Photometry Working Group together with Carme Jordi. I worked, especially with Frédéric Arenou and Jos de Bruijne, on the optimal sampling or windowing of the CCDs, i.e. the definition of the optimal window of pixels to be transmitted to ground around each star. The windowing means that about 99.9 per cent of the pixels can be skipped since most of the sky contains no stars, even when one billion stars on the sky are detected. This means lower noise in the data and much less data to be transmitted to ground. There are 64 reports since 1997 with authors from Copenhagen about sampling, detection and imaging for GAIA or Gaia.

With collaborators in Copenhagen, C. Fabricius, J. Knude, H.E.P. Lindstrøm, S. Madsen, V.V. Makarov, I.D. Novikov, A.G. Polnarev, H.J. Sørensen, and M. Vaccari, the instrument design and the possibilities to detect and measure certain objects were studied. Signatures of photometric and astrometric microlensing events, supernovae, galaxies, and NEOs were considered, cf references in ESA 2000 and papers in Leiden 1998, Les Houches 2001, and Vilnius 2001. The results were not always promising, but it is important to quantify the possibilities at an early stage where the design might still be adapted. I had learnt this lesson in 1981 when my proposal of the Tycho project, the special use of the Hipparcos star mapper, came almost too late to be implemented in the Hipparcos satellite which had been approved one year before. The numbered reports from Copenhagen on GAIA or Gaia since 1997 reached 185 when my term in the science team ended in 2007, after 32 years in ESA teams on astrometry since October 1975.

The design of optics and detection from 1998 underwent great changes in the years after 2000. The satellite had to be decreased in order to fit inside a Soyuz launcher instead of the Ariane 5 dual launch, and weight and cost problems had to be solved. The 1998 design contained two large and completely separate telescopes for astrometry, each with its own focal plane. The smaller telescope contained two focal planes, one for spectrometry and one for photometry. Several different CCDs with different size of pixels were required.

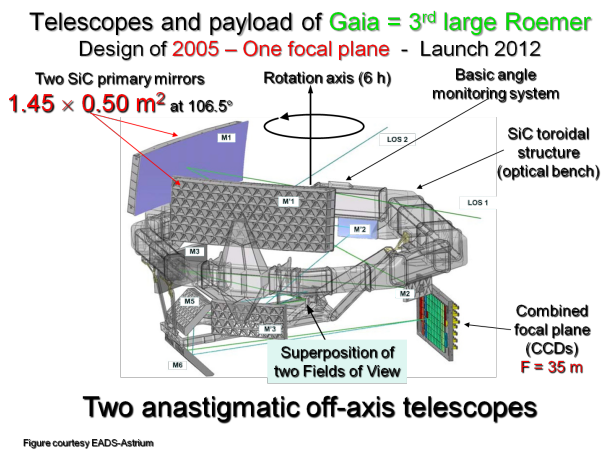

**Figure 21.** The Gaia design of 2005 by EADS-Astrium.

The final design of 2005 is very much simpler and fitting the ESA cost envelope, which is in fact the same as for the Hipparcos mission when transferred to the same economic situation. There are only two telescopes of 1.45 x 0.50 $m^2$ apertures and only one focal plane of 0.7 x 0.7 degrees holding 106 large-format CCDs, performing: star detection, astrometry, photometry and spectrometry. Only one type of CCD though with different sensitization is required. The beams from the two primary mirrors are separated by a 'basic angle' of 106.5 degrees and are brought together by a beam combiner placed at the intermediate focus. The filter photometry has been replaced by low-dispersion spectrophotometry. The penalty for astrometry has been an increased expected standard error at, e.g., V=15 mag from 11 μarcsec to about 25 which is still within the goal set originally.

## 9. The future

The present report describes the chain of ideas and actions which lead to the proposals of Roemer, GAIA and Gaia in the 1990s, especially to the evolution of the various optics and detector systems. The scientific and technical environment has been included so that the development can be properly understood. It must not be read as a history of the Gaia project, nor of course as a history of space astrometry in total, but such histories should be written. - I will end here with a view of the future.

The Gaia mission will deliver astrometric data of high accuracy, beginning a few years after launch and with final data by 2020. The astrometry will be global, covering the entire sky to $20^{th}$ magnitude with stellar distances, positions and proper motions for astrophysical and all kinds of use. The photometric data for the same one billion stars at 100 epochs during the five or six year mission will provide a unique census for study of stellar variability. This data set will be unrivalled in its kind for several decades since it is difficult to imagine that any space agency will approve a new and better mission and be ready to launch before 2040, considering for instance the great difficulties encountered at the approval of Hipparcos in 1980 and of the Cornerstone study in 1997.

On such time scales, it is hard to imagine that the astrometric expertise in the present Gaia teams can be preserved for a new mission. This was much easier for Gaia because a realistic design of a new mission was available already in 1992 while Hipparcos was still operating. This was possible since CCD detectors were highly developed at that time and it could be seen that they would be much more efficient in an astrometric satellite than the photoelectric detectors used in Hipparcos. No similar technological basis for a great improvement is known to me, but a new mission similar to Gaia should be considered.

All-sky scanning satellites as Hipparcos and Gaia cannot stop and stare at selected stars or areas. This requires pointing satellites and such missions for limited sky coverage and a smaller number of stars, but with higher accuracy and/or at redder wavelengths may have a better chance of approval.

Higher accuracy can in principle be obtained by more photons (i.e. longer integration times and/or larger apertures) and/or higher angular resolution, for instance by interferometry. It is difficult to imagine that NASA or any other space agency will soon engage in interferometry for astrometry after the long effort on SIM has been stopped, SIM 2011, and after the lesson from the study of the interferometric version of GAIA 1993-97.

The long-term Japanese plans for high-accuracy infrared astrometry with Jasmine of the Galactic bulge look promising, Gouda et al. 2009 and Jasmine 2011. A small low-cost scanning satellite called "Nano-JASMINE" is due for launch in August 2011. A "Small Jasmine" and "Jasmine" are both pointed satellites and launches are expected in 2016 and in the 2020s.

The J-MAPS astrometric mission by the USNO is a pointing satellite using hybrid CMOS/CCD detectors, but it is an all sky mission using the overlapping plate solution method. It is expected to be launched in 2014 and it will observe stars to 14$^{th}$ magnitude, Hennessy & Gaume 2009.

**Acknowledgements:** I would like to thank Timo Prusti and his collaborators for the Gaia schedule *without Roemer* and Francois Mignard for his slide with Roemer as one of the *"unsuccessful followers"* of Hipparcos because they thereby got me started on the present report, and I thank Francois and Timo for their warm welcome at my subsequent lectures on this subject at respectively CNES in Toulouse and at ESTEC. I am grateful to Ulrich Bastian, Lars Brink, Anthony Brown, Tom Corbin, Aase Høg, Carme Jordi, Lennart Lindegren, Francois Mignard, Michael Perryman, Rudolf Le Poole, Ken Seidelmann, and Sperello di Serego Alighieri for correspondence during the recent months about the subject and for comments to previous versions of this report.

## 10. References on space astrometry

The following list was compiled on the development of space astrometry especially in the 1990s, but more references are included than mentioned in the present report *though without being complete.*

The references are given in the sequence they were presented publicly at conferences or by other distribution. The year of publication in proceedings is often a year later. Reports on Hipparcos and Tycho at the symposia are mostly omitted.

Morrison and G.F. Gilmore (eds.) *Galactic and Solar System Optical Astrometry*. Proceedings of a conference held in Cambridge June 21-24, 1993. 246-252.

**To ESA in October 1993:**
Lindegren L., Perryman M.A.C., Bastian U., Dainty J.C., Høg E., van Leeuwen F., Kovalevsky J., Labeyrie A., Mignard F., Noordam J.E., Le Poole R.S., Thejll P., Vakili F. 1993b. GAIA: Global Astrometric Interferometer for Astrophysics, proposal for a Cornerstone Mission concept submitted to ESA on 12 October 1993, including a cover letter by Lindegren, 6 pages. Available at www.astro.ku.dk/~erik/gaia_proposal_1993.pdf

**March 10, 1994:**
Seidelmann writes to me that USNO is pursuing NEWCOMB which is however unfunded, and that they are interested in collaboration on other space-based astrometry projects like Roemer and GAIA.

**April 1994:**
Lindegren L. 1994. Letter in April 1994 to Steven Beckwith, chairman of the UV-to-radio topical team of the Survey Committee then drafting ESA's Horizon 2000+ plan, 1+7 pages. Available at www.astro.ku.dk/~erik/beckwith_lindegren.pdf

**The Hague in August 1994:**
Høg E. & Seidelmann P.K. (eds.) 1995, *Astronomical and Astrophysical Objectives of Sub-milliarcsecond Optical Astrometry,* Proceedings of the 166th Symposium of the IAU, held in The Hague, The Netherlands, August 15-19, 1994,16+441 pp.

Høg E. 1995a, A New Era of Global Astrometry. II: A 10 Microarcsecond Mission. (Including proposal of Roemer+). In: Proceedings of the 166th Symposium of the IAU, 317.

Johnston K., Seidelmann P.K., Reasenberg R.D., Babcock R., Philips J.D. 1995, Newcomb Astrometric Satellite. In: Proceedings of the 166th Symposium of the IAU, 331.

Lindegren L. and Perryman M.A.C. 1995, A Small Interferometer in Space for Global Astrometry: The GAIA Concept. In: Proc. of the 166th Symposium of the IAU, 337.

**To ESA in September 1994:**
Lindegren L. & Perryman M.A.C. 1994. GAIA: Global Astrometric Interferometer for Astrophysics (A Concept for an ESA Cornerstone Mission). Supplementary Information Submitted to the Horizon 2000+ Survey Committee.

**Cambridge in June 1995:**
Perryman M.A.C. & van Leeuwen F., eds. 1995, Proceedings of a Joint RGO-ESO Workshop on "Future Possibilities for Astrometry in Space", Cambridge, UK, 19-21 June 1995 (**ESA SP-379**, September 1995), 323 pp.

Some of the following reports are available at www.astro.ku.dk/~erik/papers/

Cecconi M. & Lattanzi M.G. 1995, GAIA Optics: Polychromatic PSF and Pupils. In: ESA SP-379, September 1995, 251.

Daigne G. 1995, Direct Fringe Detection and Sampling of the Diffraction Pattern. In: ESA SP-379, September 1995, 209.

Fabricius C. & Høg E. 1995, Scientific Analysis of Data from a Scanning Satellite. In: ESA SP-379, September 1995, 273.

Favata F. & Perryman M. 1995, Parallel Aquisition of Radial Velocities and Metallicities for a GAIA-Type Mission. In: (ESA SP-379, September 1995, 153.

Gai M., Lattanzi M.G., Casertano S. & Guarnieri M.D. 1995, Non-Conventional Detector Applications for Direct Focal Plane Coverage. In: ESA SP-379, September 1995, 231.

Gilmore G. & Høg E. 1995, Key Questions in Galactic Structure with Astrometric answers. In: ESA SP-379, September 1995, 95.

Høg E. 1995b, Observation of Microlensing by an Astrometric Satellite. In: ESA SP-379, September 1995, 125.

Høg E. 1995c, Some Designs of the GAIA Detector System. In: ESA SP-379, September 1995, 223.

Høg E. 1995d, Three Astrometric Mission Options and a Photometric System. In: ESA SP-379, September 1995, 255.

Høg E. 1995e, Astrometric Satellite with Sunshield. In: ESA SP-379, September 1995, 263.

Lindegren L. 1995, Summary of the Parallel Sessions. In: ESA SP-379, September 1995, 267.

Lindegren L. & Perryman M.A.C. 1995, The GAIA Concept. In: ESA SP-379, September 1995, 23.

Loiseau S. & Shaklan S. 1995, Analysis of an Astrometric Fizeau Interferometer for GAIA. In: ESA SP-379, September 1995, 241.

Makarov V. 1995c, Gravitational Waves as Astrometric Targets. In: ESA SP-379, September 1995, 117.

Noordam J.E. 1995, On the Advantage of Dispersed Fringes. In: ESA SP-379, September 1995, 213.

Perryman M.A.C. & Peacock A. 1995,A Superconducting Detector for a Future Space Astrometry Mission. In: ESA SP-379, September 1995, 207.

Rabbia Y. 1995, Obtaining Spectral Information from an Astrometric Interferometric Mission. In: ESA SP-379, September 1995, 217.

Seidelmann P.K. et al. 1995, A Fizeau Optical Interferometer Astrometric Satellite. In: ESA SP-379, September 1995, 187.

Straizys V. & Høg E. 1995, An Optimum Eight-Colour Photometric System for a Survey Satellite. In: ESA SP-379, September 1995, 191.

Yershov V.N. 1995, An Ultraviolet Option for a Future Astrometric Satellite. In: ESA SP-379, September 1995, 197.

**FAME-1 in 1995:**
Johnston K. 1995a June, Step 1 Proposal to NASA for the Medium-class Explorer (MIDEX), Fizeau Astrometric Mapping Explorer (FAME). Technical Report with 40 pages.

# En landmåler i himlen

*Af Erik Høg, Niels Bohr Institutet*

> Erindringer om 50 år med astrometrien, der begyndte ved en høstak syd for Holbæk og førte til bygning af to satellitter. Et videnskabeligt højdepunkt er stjernekataloget Tycho-2, der nu er helt uundværligt ved styring af satellitter og ved astronomiske observationer.

Der var engang et Astronomisk Observatorium ved Københavns Universitet, men det er der ikke mere. Observatoriet havde endda to afdelinger, en på Øster­voldgade i en smuk bygning fra 1861, en anden fra 1953 ved den lille landsby Brorfelde syd for Holbæk, hvor der var teleskoper, værksteder og et aktivt viden­skabeligt liv. Men nu om dage foretages astronomiske observationer fra bjergtoppe i Chile og på La Palma og fra satellitter, så for 14 år siden flyttede afdelingen fra Brorfelde sammen med afdelingen fra Østervold til en bygning nær Rigshospitalet, som også huser geofysikere og rumforskere.

Siden er hele astronomien blevet lagt ind i Niels Bohr Institutet, så der slet ikke mere er noget, der hedder Astronomisk Observatorium ved Københavns Universitet. Hermed slutter en epoke, der begyndte i 1642 med indvielsen af Observatoriet på Rundetårn. Vores mest berømte astronom, Tycho Brahe (1546-1601), hørte direkte under kongen og havde ingen tilknytning til Universitetet, der dengang var temmelig forbenet, mens Tychos observatorium på Hven var det mest moderne, der overhovedet fandtes, hvor man målte stjerners og planeters positioner med hidtil uset nøjagtighed.

Astronomi er fysik, siger man, og det er meget rigtigt. Jeg klager ikke over udviklingen. Det kunne også lige passe, når jeg siden 1975 har været med til at finde på to satellitter, Hipparcos og Gaia, og været med i både udvikling og anvendelse, takket være Danmarks medlemskab af ESA, det Europæiske Rumagentur. Det drejer sig om to satellitter, der er specielt konstrueret til at måle stjerners positioner meget nøjagtigere, end nogen tidligere har gjort. Amerikanere og russere har prøvet, men de har endnu ikke kunnet bygge en satellit til astrometri, det kan kun ESA.

## Astrometri

Astrometri er den gren af astronomien, som netop Tycho Brahe dyrkede ved at måle stjerners positioner. Ved målinger gennem nogle år får man også stjernernes bevægelser og afstande. Resultaterne anvendes i alle grene af astronomien for at få en fysisk forståelse af hele Universets opbygning af stjerner som glødende gaskugler, af planeter, der kredser om Solen og omkring andre stjerner, af galakser som roterende stjernesyste­mer osv. osv.

Danske astronomer har leveret to særlige bidrag til astrometrien i de sidste halvtreds år, begge med udgangspunkt i den nye meridiankreds, der blev op­stillet i Brorfelde i 1953 og begge baseret på den fotoelektriske teknik til astrometri, som jeg udviklede i mine femten år ved observatoriet i Hamborg fra 1958. Det ene astrometriske bidrag består i udvikling af en automatisk meridiankreds og observationer med denne fra 1984 i 2000 meters højde på La Palma, den vestligste af de Canariske Øer. Det andet bidrag begynder i 1975 og angår udvikling og anvendelse af den første astrometriske satellit, Hipparcos, som det følgende handler om.

En meridiankreds er en kikkert opstillet på en særlig måde, som blev opfundet af Ole Rømer for 300 år siden, og meridiankredse var længe de nøjagtigste instru­menter til måling af stjerners positioner, lige indtil de blev udkonkurreret af Hipparcos-satellitten. Kikkerten sidder vinkelret på en akse, der hviler på en søjle i øst og en anden søjle i vest, således at man kun kan se stjerner, når de passerer nord-syd retningen. Observatøren ser en stjerne komme ind i synsfeltet og glide igennem på grund af Jordens omdrejning. Han måler det nøjagtige tidspunkt for stjernens passage af meridianen, og han måler kikkertens hældning med vandret ved aflæsning af en nøjagtig delekreds, som er fastgjort på aksen. Disse to målinger giver stjernens position svarende til et steds længde og bredde på Jorden.

Dette var princippet, og i praksis er mange metoder udviklet for at opnå den størst mulige nøjagtighed. Indtil 1950 var det altid en person, der virkelig ob­serverede stjernen under passagen, så forsøgte man en fotografisk metode for eksempel i Brorfelde, men fra 1960 udviklede jeg en automatisk fotoelektrisk metode, hvorved de nødvendige målinger blev skrevet på en hulstrimmel, der derefter kunne læses ind i en computer.

## Fra høstak til satellitter – at gøre nytte var det, jeg altid ville

Astrometriens udvikling med Hipparcos-satellitten betød en revolution, som man kun kan forstå om­fanget af, hvis man kender den videnskabelige bag­grund. Astrometri var hovedsagen i astronomien i århundreder, og en meridiankreds var hovedinstrument i ethvert observatorium i 1800-tallet. Det ændredes med fysikkens udvikling, især med atomfysikken i 1900-tallet, og astrometri blev efterhånden betragtet som ganske vist uundværlig, men dog som kedelig og besværlig, ganske som Peter Naur så drastisk udtrykker det nedenfor. Så de fleste observatorier interesserede sig efterhånden kun for astrofysik. Men der var undtagelser som København, hvor den fremragende astrofysiker

Bengt Strömgren kendte astrometriens værdi og bestilte en ny meridiankreds til sit nye observatorium på en 90 meter høj bakke ved Brorfelde syd for Holbæk.

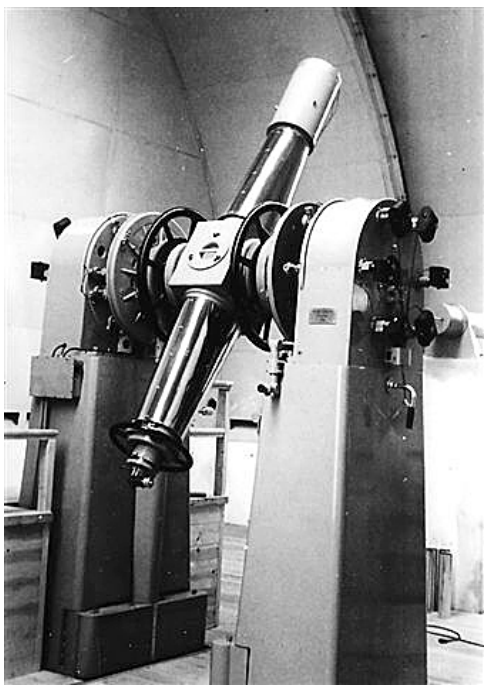

**Figur 1.** Meridiankredsen i Brorfelde i 1954, som jeg arbejdede med som student.

Peter Naur arbejdede fra 1955 på det nye observatorium i Brorfelde. Men han var dybt frustreret over arbejdet, som bare gik ud på at udvikle den nye meridiankreds og så lave landmåling på himlen. "Der var ikke for fem flade øre videnskabelighed i det, det var ligesom landmålerarbejde! Der var jo ingenting i det! Det eneste der var, var instrumentudvikling og derefter var det fabriksarbejde!" Sådan siger han i et interview med en historiker [1] halvtreds år efter og bliver derved ganske animeret og vred. Han forlod derfor observatoriet få år efter og blev snart Danmarks første professor i datalogi. I mange år var han den, der altid blev spurgt af radio og aviser om computere. Han var en sand pressens guru dengang.

Jeg var den første, der observerede i Brorfelde, allerede fra august 1954, da jeg var 22 år. Jeg var helt alene derude, og sov til tider i en høstak efter observationerne. Jeg var student hos Naur og lærte meget af ham. Vi havde et udmærket samarbejde i de år med udvikling af meridiankredsen, men jeg havde et helt andet syn på dette arbejde end Naur. Det var noget, der lige lå for mig, jeg havde interesse for udvikling af instrumenter, og jeg vidste, at arbejdet var vigtigt for astronomiens udvikling, og at gøre nytte var det, jeg altid ville. Jeg havde bygget kikkerter som dreng, da jeg var 16 år og selv slebet spejlene.

Min interesse og evner for udvikling af instrumenter førte i det lange løb til opsendelse af satellitten Hipparcos, der observerede stjernernes positioner, bevægelser og afstande meget nøjagtigere end nogen før havde gjort. Dertil hører, at jeg i femten år fra 1958 arbejdede ved observatoriet i Hamborg.

**Jeg vil tælle fotonerne, lysets mindste dele**

Ved observatoriet i Hamborg havde man planer om en ekspedition til Perth i Vestaustralien med en gammel meridiankreds. Men jeg foreslog i 1960 at tælle fotonerne til måling af lyset og at bruge en computer til beregningerne. Man lader stjernen glide hen over nogle smalle spalter, mens man hele tiden måler det lys, der kommer igennem spalterne. Derved måles både stjernens position og dens lysstyrke. Teknikken til elektronisk tælling kendte jeg før andre astronomer, fordi jeg havde mødt den som soldat, idet vi havde målt radioaktivitet i støv fra atmosfæren, som stammede fra stormagternes sprængning af brint- og atombomber.

Direktøren Otto Heckmann var straks med på mine idéer om fotoelektrisk astrometri, og han havde tillid til, at jeg ville holde ud i alle de år, der lå foran med udviklingen, som faktisk tog syv år. Vi købte den bedste computer til formålet, en regnemaskine hed det dengang, og den bedste maskine var den danske GIER, som også kom med til Australien. Det var en stor astronomisk ekspedition efter den tids målestok, idet en stab på seks til ti mennesker foretog observationer og beregninger gennem fem år. Resultaterne for 25.000 stjerner blev udgivet i 1976 af E. Høg og J. von der Heide, idet Leif Helmer udførte de sidste beregninger af kataloget, efter at jeg var kommet tilbage til Brorfelde.

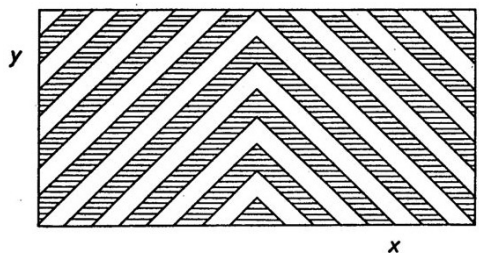

**Figur 2.** Gitter til fotoelektrisk astrometri, Mit første forslag i 1960, som franske astronomer kaldte 'une grille de Høg'.

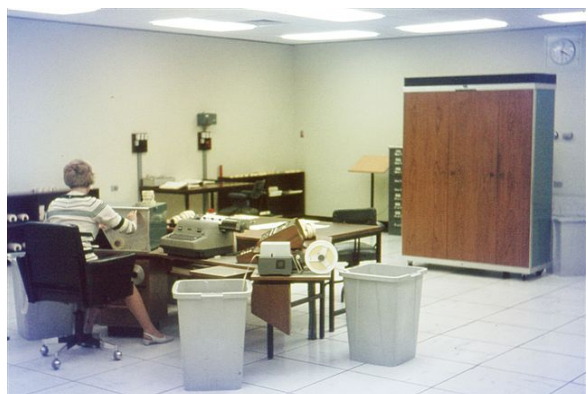

**Figur 3.** GIER i Australien 1970.

I mellemtiden var udviklingen i Brorfelde gået videre, idet man havde bygget et fotografisk kamera til observation af stjernen. Man indså dog snart, at den fotoelektriske metode, som samtidig blev udviklet i Hamborg, var langt mere effektiv, idet alle observationer kom direkte på hulstrimler, der senere kunne stikkes i regnemaskinen. Derfor blev meridiankredsen i Brorfelde udrustet med fotoelektrisk teknik, da jeg kom tilbage til Brorfelde i 1973, og indstilling på stjernen blev efterhånden automatiseret. Derefter blev meridiankredsen i 1984 flyttet til La Palma, der har mange stjerneklare nætter, og hvor man i mange år udførte astrometriske målinger, ledet af Leif Helmer i samarbejde med astronomer fra England og Spanien. Imidlertid var dette en parallel udvikling, der ikke havde nogen indflydelse på den udvikling af astrometri fra en satellit, som jeg nu skal beskrive.

**Jeg vil bygge en satellit**

Jeg skrev artikler og holdt mange foredrag om mine erfaringer med fotoelektrisk astrometri. Så det var meget naturligt, at ESA i 1975 spurgte mig, om jeg ville være med i overvejelserne om en satellit til astrometri. Den franske astronom Pierre Lacroute havde i mange år arbejdet med denne store vision og med planer, som imidlertid forekom mig ganske urealistiske. Jeg havde derfor ingen interesse på forhånd, men jeg måtte jo sige ja til at være med. Allerede ved det første møde i Paris blev jeg faktisk entusiast på sagen, fordi formanden sagde, at vi slet ikke skulle tænke på Lacroutes planer, men kun på hvordan vi bedst kunne udnytte rumfartens tekniske muligheder i vores videnskab. Så kunne jeg pludselig tænke helt frit, og selvom jeg ellers aldrig havde interesseret mig for rumteknologi, lavede jeg på seks uger et nyt design af en astrometrisk satellit.

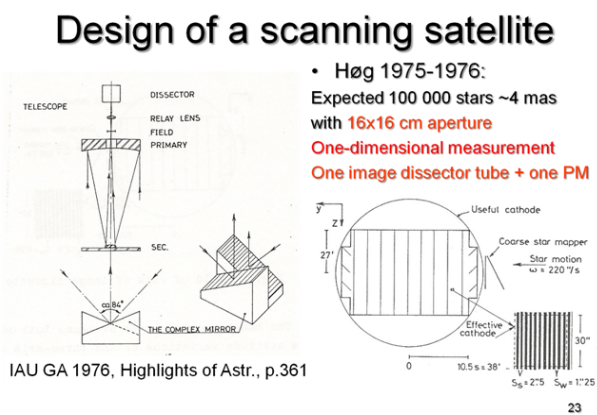

**Figur 4.** Mit design i 1976 af Hipparcos, kikkert og fokalplan med detektorer og forklaring i billedet.

Mit design var helt nyt, men det slog hurtigt igennem. Samarbejdet med astronomer fra alle steder i Europa og med ingeniørerne hos ESA og i industrien var lige noget for mig. Satellitten Hipparcos blev vedtaget i 1980 og opsendt med en Ariane-raket i 1989. Resultaterne blev trykt i 1997 og har siden revolutioneret astronomien på mange områder, hvad der dog ikke er plads til at beskrive her, se f.eks. [2].

Tre år senere kunne vi udgive et katalog kaldet Tycho-2 med 2,5 millioner stjerner. Det er blevet det mest anvendte katalog over himlens klare stjerner til brug i astronomien og ved styring af satellitters bevægelse og drejninger i rummet. Navnet Tycho havde jeg valgt efter Tycho Brahe, da jeg i 1981 fandt på, at man skulle lave nogle ekstra målinger med Hipparcos-satellitten. Mit forslag kom altså et år efter satellittens vedtagelse, og på det tidspunkt vil man normalt altid sige, at der er du altså bare for sent ude. Men mit forslag vakte sådan tilslutning, at de ekstra millioner blev bevilget, så de nødvendige ændringer i designet kunne indbygges. Også navnet Tycho gik igennem, og det var jeg meget glad for. Jeg var nemlig blevet lidt snydt med navnet på satellitten, som jeg foreslog i 1975. Jeg havde dengang kaldt den Tycho, men det blev senere ændret til Hipparcos på en måde, jeg ikke skal komme ind på. Navnet Hipparcos skal erindre om den græske filosof Hipparchos (190-120 fvt.), der også kaldes astronomiens fader.

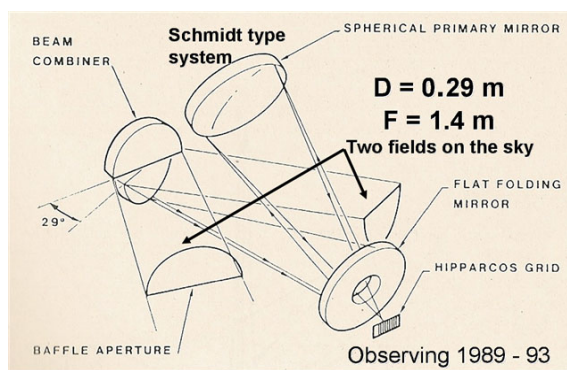

**Figur 5.** Hipparcos' optik i endelig form [3].

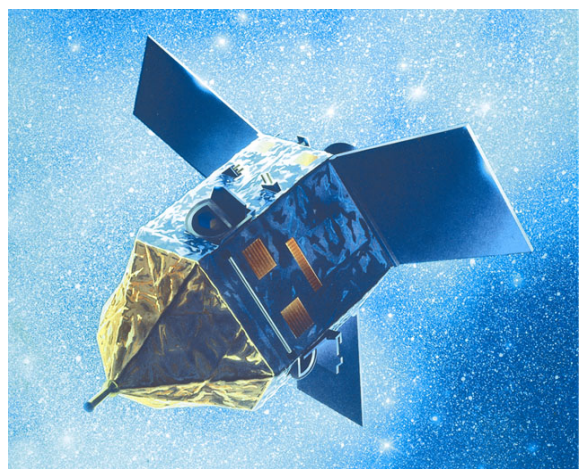

**Figur 6.** Hipparcos-satellitten.

**Hipparcos kom i en forkert bane om Jorden**

Opsendelsen i august 1989 var en stor begivenhed for os alle efter mange års indsats, for mit eget vedkommende siden 1975. Det foregik i Sydamerika i fransk Guyana, hvorfra de fleste ESA-satellitter opsendes, og under stor bevågenhed med mange hundrede tilskuere. Nogle af os var udvalgt til at flyve derover i en Concorde fra Paris over Dakar i Vestafrika. Det var ret specielt, at Solen ikke ville gå ned, mens vi fløj mod vest sent på eftermiddagen med 2200 km i timen eller mere end to gange lydens hastighed.

Jeg var blandt de begunstigede også under selve opsendelsen, fordi jeg var leder af to konsortier af astronomer, der havde påtaget sig at udføre beregningerne på alle observationer fra satellitten. Det ene var et Hipparcos-konsortium, der skulle beregne meget nøjagtige positioner, bevægelser og afstande for de 120.000 stjerner, der udgjorde Hipparcos-missionens vigtigste opgave. Det andet var Tycho-konsortiet, der skulle behandle alle Tycho-observationerne for at nå frem til positioner for mindst 400.000 stjerner med noget mindre nøjagtighed, men dog bedre end man kunne måle fra Jorden. Det blev i virkeligheden til en million stjerner med beregningerne for Tycho indtil 1996, og derefter udførte vi nye beregninger med bedre computere, og det blev i år 2000 til Tycho-2 kataloget med de 2,5 millioner lysstærkeste stjerner på himlen. Der var endnu et Hipparcos-konsortium, ledet af Jean Kovalevsky med samme opgave for de 120.000 stjerner for at sikre en uafhængig kontrol af den enorme og enormt vigtige opgave – og det lykkedes for os alle gennem tyve års samarbejde. De 120.000 stjerner var blevet udvalgt af et særligt konsortium på basis af flere hundrede forslag fra astronomer, og det var i sig et kæmpearbejde, der blev ledet af Catherine Turon.

Men nu tilbage til opsendelse af Hipparcos i august 1989. Vi sad og ventede kun to km fra den 58 meter høje Ariane raket. Vi var blevet instrueret om en eventuel ulykke. Hvis hele raketten eksploderede, var vi i fare for forgiftning af den røg, der kunne nå hen til os. Hvis alarmen lød, skulle vi skynde os hen til vort køretøj og straks tage iltmaskerne på. Imidlertid så vi motorerne tænde, røgen vældede op, og raketten begyndte at stige, men til min forundring hørte vi intet! Først efter seks sekunder nåede den stærke brølen hen til os – for lyden tager jo tid.

Vi kunne nu følge raketten på vej op, og efterhånden så vi kun ilden fra raketten, mens jeg stod og snakkede med mine kolleger Turon, Kovalevsky og andre. Vi troede at alt gik godt, idet både andet og tredje trin af raketten havde tændt, som de skulle.

Da jeg var tilbage på hotellet, prøvede jeg at ringe til min gamle mor i Danmark, og det viste sig at være meget besværligt, fordi det skulle gå over den franske ø Martinique. Mor ville sikkert blive glad for en opringning fra Sydamerika, det havde hun aldrig fået før, og hun blev glad. Da jeg lagde røret, var det som en spænding udløstes, for pludselig stod tårerne ud af hovedet på mig, af glæde selvfølgelig.

Men to dage senere hørte vi om en meget alvorlig fejl. Satellitten blev først skudt ind i en langstrakt elliptisk bane, hvor den ene ende er et par hundrede kilometer over Jordens overflade, mens den anden ende er 36.000 km oppe. I denne bane tager et omløb cirka 11 timer, og der bliver den i nogle omløb, mens man afprøver, at alt er i orden. Når man er klar, og satelitten er i det højeste punkt, tænder man en raket, som sidder inde i satellitten og har en ladning brændstof på 500 kg. Det skal brænde i nogle minutter for at give satellitten en større hastighed, så den kommer ind i en cirkulær bane om Jorden, hvor et omløb tager 24 timer. Så vil satellitten til stadighed stå lige over et punkt på Jordens ækvator, idet Jorden også drejer en omgang på 24 timer. Det betyder, at radioforbindelsen hele tiden kan varetages af en eneste parabolantenne på Jorden, som står et sted i nærheden af Darmstadt, hvor ESAs kontrolcenter ligger.

Sådan skal det normalt ske, men denne gang ville raketmotoren ikke tænde, så banen blev ved med at være elliptisk, og det var en katastrofe. Satellitten vil fire gange i døgnet passere nogle strålingsbælter, som indeholder partikler, der hurtigt vil ødelægge elektronikken, og man kan ikke længere nå den med antennen ved Darmstadt. Sådan var situationen, da jeg nåede hjem til Danmark. En ulykke er jo altid en god nyhed for pressen, så jeg var gæst i aftennyhederne i begge fjernsynskanaler. Jeg sagde, at Hipparcos nok kun kunne give observationer i nogle måneder, og at de allerede ville være en stor gevinst for astronomien, men vi måtte have en ny satellit.

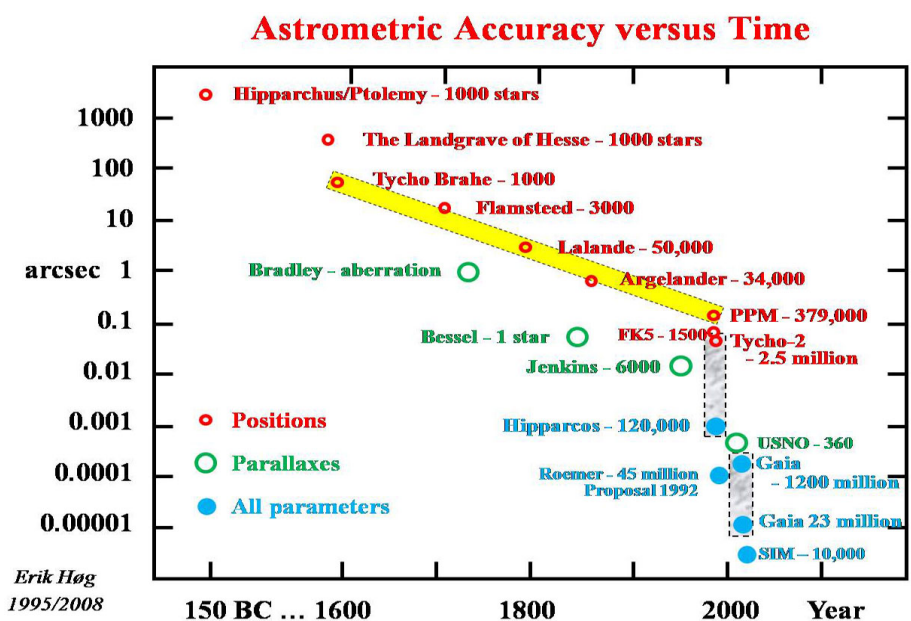

**Figur 7.** Astrometrisk målenøjagtighed gennem 2000 år. Tycho Brahe formindskede målefejlene til en femtedel. Derefter gik det gradvis i 400 år, indtil Hipparcos gjorde et spring på en faktor 100, og Gaia vil fortsætte.

Da jeg var til en konference i Leningrad i september og holdt foredrag om Hipparcos og Tycho, var der enorm interesse blandt astronomerne og stor medfølelse med os. En distingveret russisk herre blev præsenteret for mig som kommende fra videnskabernes akademi og fra russisk rumfart, og han tilbød at sende en ny satellit op med en russisk raket. I ESA var der selvfølgelig stor aktivitet for at redde missionen, og det lykkedes virkelig at få gode observationer i tre år, det tidsrum der oprindeligt var planlagt. Nøjagtigheden blev faktisk meget bedre, end vi havde regnet med på forhånd, og var vi kommet i den rigtige bane, havde vi sikkert kunnet få gode observationer i fem år i stedet for tre, men i historisk perspektiv ville det ikke have gjort den store forskel. Hipparcos blev også under de givne omstændigheder en milepæl i astronomien historie.

Elektronikken blev beskadiget af partiklerne i strålingsbælterne, men meget langsommere end frygtet, og vore data blev hentet ned med fire radioantenner efter tur på Hipparcos' vej omkring Jorden. Jeg er blevet spurgt, hvordan jeg følte det, om jeg havde nedture under de største genvordigheder. Men jeg havde nok at tage mig til, så jeg var altid ved godt mod, det ligger simpelthen til mig. Du skal leve hver dag, som om det er den vigtigste i dit liv. Ikke den sidste dag, men den vigtigste.

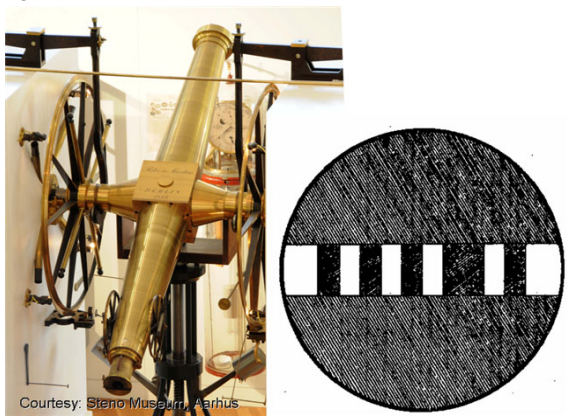

**Figur 8.** Meridiankredsen i København, som Bengt Strömgren arbejdede med i 1925. Til højre ses de spalter, som stjernen gled hen over.

## Om seks astronomer og om at have heldet med sig

Udviklingen indtil vedtagelsen af Hipparcos i 1980 var meget afhængig af ganske få astronomer. Denne udvikling begyndte faktisk i 1925, med at Bengt Strömgren lavede nogle eksperimenter på den gamle meridiankreds i København. Han var teenager, bare 17 år, da han påviste, at stjerners positioner kan måles, når man lader en stjerne glide hen over nogle spalter. Man opfanger stjernens lys med en fotocelle, der omsætter lyset til en elektrisk strøm. Strömgrens fotoelektriske teknik var den bedste på hans tid, men alligevel så primitiv, at den ikke duede i praksis. I 1960 kunne jeg omsætte Strömgrens ide til en tælling af fotonerne, og det virkede fint på meridiankredse og senere også i Hipparcos-satellitten.

Seks astronomer spillede hver deres helt afgørende rolle i udviklingen. Bare een af dem havde manglet, ville denne udvikling være gået i stå. Der ville ikke være blevet vedtaget nogen astrometrisk satellit i 1980, og formentlig aldrig. Disse kritiske år fra 1925 til 80 har jeg skrevet om i en række historiske artikler, som bl.a. ligger på min hjemmeside. To af de seks astronomer var danske, Bengt Strömgren og jeg selv.

Når jeg selv kom til at spille en rolle i disse kritiske år, skyldes det selvfølgelig både talent og held. Hvor stor en rolle heldet har spillet for mig, vil jeg gerne have lov at fortælle. I 1958 tog jeg på et ophold i Hamborg, som var planlagt til at vare 10 måneder, og det var især Peter Naur, der tilskyndede mig til at rejse. Jeg selv syntes ellers ikke jeg var dygtig nok endnu, men Naur sagde, at jeg jo netop skulle rejse for at blive dygtigere. Hamborg var en af mulighederne, men ikke på grund af den gamle meridiankreds, der dog kom til at spille en meget stor rolle. Nej, jeg ville arbejde med astrofysik, og begyndte med spektre, som vi optog med observatoriets store Schmidt-kikkert. Det gik så godt, at opholdet blev forlænget ud over de første 10 måneder. Så skete det i 1960, at jeg fik den gode ide om fotoelektrisk astrometri, som passede i planerne om ekspeditionen til Australien. Det var et held, og det var et held, at direktøren troede på mig og disse nye idéer. Det var bestemt ikke nogen selvfølge, så Otto Heckman er en af de seks astronomer. Pierre Lacroute og Jean Kovalevsky, der altid støttede Lacroute, er to franske blandt de seks astronomer. Den sjette må nævnes her, Lennart Lindegren fra Sverige, langt den yngste af os alle, født 1950, og uden hvem vi ikke kunne have mestret dataanalysen for Hipparcos lige fra starten. Uden Lindegrens tidlige indsats ville Hipparcos-satellitten ikke være blevet vedtaget i 1980 i den hårde konkurrence med de astrofysiske projekter.

Lindegren var ung student, da jeg i 1973 første gang mødte ham i Lund, hvor han arbejdede med en gammel meridiankreds. Mine kolleger sagde senere rosende, at jeg "fandt Lennart". Jeg fik gjort ham interesseret i nogle observationer af planeter fra meridiankredsen i Perth. Han lavede en fremragende analyse af målingerne og fik sin grad fra universitetet i 1976. Jeg havde så et møde med ham, hvor jeg fortalte om mine planer for en satellit og beskrev den enorme regneopgave, som bestod i at løse ti millioner ligninger med en halv million ubekendte. Kun fire uger senere sendte Lennart mig en nøjagtig beskrivelse på ti sider af den matematiske metode, som vi faktisk udviklede og anvendte på de virkelige observationer fra Hipparcos. Det blev til den største samlede beregning af observationer i astronomiens historie. Lindegren har fortsat i samme tempo lige siden, og i 1992 overtog han rollen fra mig som leder af det ene Hipparcos-konsortium.

Hvis jeg var blevet i Danmark i 1958 eller var kommet tilbage efter et ophold til Hamborg, ville jeg have haft et stadigt problem med den nye direktør, professor Anders Reiz, som tiltrådte i 1958 som Bengt Strömgrens efterfølger. Reiz og jeg havde forskellig kemi, det mærkede jeg i alle vore samtaler i mange år. Han var ikke positiv over for mine nye idéer, men han kom til at betyde meget for dansk astronomi ved at fostre en hel generation af astrofysikere. Reiz var også astrometriker og havde arbejdet med meridiankredsen i Lund, så han var bestemt ekspert på området. Han var helt opsat på at udvikle den fotografiske metode til observation af stjernen med meridiankredsen, en udvikling der var begyndt i Lund, og som han var enig med Strömgren i. Hvis jeg var kommet med min fotoelektrisk metode, ville Reiz aldrig have kunnet gå ind på den, også fordi han var nødt til hurtigst muligt at få instrumentet i Brorfelde, observatoriets hovedinstrument, til at producere observationer. Det lykkedes for Svend Laustsen i begyndelsen af tresserne. Svend sagde for nylig til mig, at han godt kunne se dengang, at den fotoelektriske metode til observation af stjernen, som jeg samtidig udviklede i Hamburg, måtte være fremtiden, men på det tidspunkt var han naturligvis nødt til at fortsætte med den fotografiske metode. Men han udviklede en metode til registrering af delekredsen på hulstrimler, ganske som vi også gjorde i Hamborg, og det var et kæmpe fremskridt.

Da jeg kom tilbage til Brorfelde i 1973, var professorvældet forbi i Danmark, og jeg kunne nu se frem til, at man ville følge mine idéer. Faktisk blev jeg hentet hjem fra Hamborg, hvor jeg på det tidspunkt for længst havde fået en livstidsstilling som tysk tjenestemand, og hvor jeg havde hus og hjem med en dansk kone og tre børn. På Hamborg Observatoriet var betingelserne for astrometri blevet meget ugunstige fra ledelsens side, så det var mit held, at jeg blev hentet væk på det tidspunkt. I Brorfelde var der et velfungerende videnskabeligt og teknisk miljø, skabt ved Anders Reiz' talent for at udnytte de "gyldne tressere" til en stor udvidelse af hele observatoriet. Her må jeg, især hvad det tekniske angår, nævne lederen af det mekaniske værksted, Poul Bechmann, og lederen af det elektroniske værksted, Ralph Florentin Nielsen, som begge kom til at spille en meget stor rolle for mit arbejde. Bechmann kunne designe mekanik meget bedre end jeg havde kendt i Hamborg, og det var en vigtig forudsætning for mit samarbejde med ham om et nyt fotoelektrisk mikrometer, mere raffineret end det vi havde sendt til Perth, og en forudsætning for den efterfølgende success med den automatiske meridiankreds på La Palma, som især Leif Helmer stod for, idet jeg havde andre jern i ilden. Florentin var den, der hjalp mig for eksempel i 1975, da jeg lavede det nye design af en satellit, som blev til Hipparcos, og i 1992, da jeg skulle lære om CCD-detektorer for at designe den nye Roemer-satellit, der blev til Gaia.

## Ny satellit i 2012

Udviklingen af astrometrien er ikke gået i stå efter Hipparcos, idet ESA bygger en ny satellit med navnet Gaia, der skal opsendes i 2012, og som er en million gange bedre end Hipparcos. Jeg har i september 2010 besøgt Gaia-satellitten i den fabrik i Toulouse, hvor den bygges sammen, og jeg holdt et historisk foredrag for ingeniørerne om udviklingen af astrometri fra rummet. Dertil hører eventyret om Roemer/Gaia-projektet, men den historie skal vente til en anden gang.

Min gamle lærer og kollega fra tiden i Brorfelde, Peter Naur, hørte for nylig mit foredrag om alt dette og sagde bagefter: "Det er utroligt, hvad du har nået, Høg". Jeg skylder Københavns Universitet tak for en god uddannelse, og Astronomisk Observatorium gav gennem årene husly til min forskning, der blev støttet af mange offentlige og private midler. Mit videnskabelige arbejde er blevet belønnet med en medalje fra ESAs direktør for videnskab og med en medalje fra det russiske videnskabernes akademi, og den Internationale Astronomiske Union har givet en asteroide navnet *ErikHøg*.

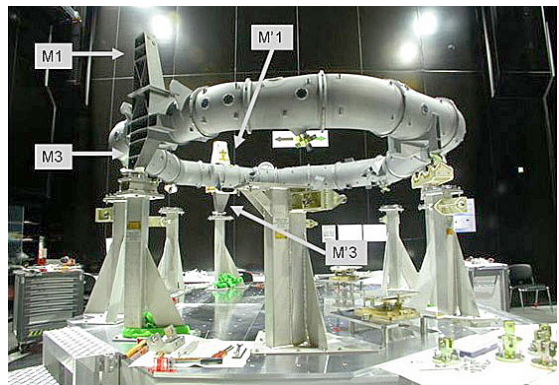

**Figur 9.** Denne torus på tre meter diameter skal bære al optikken i Gaia. Materialet er siliciumcarbid (SiC), der har en termisk udvidelseskoefficient på nul.

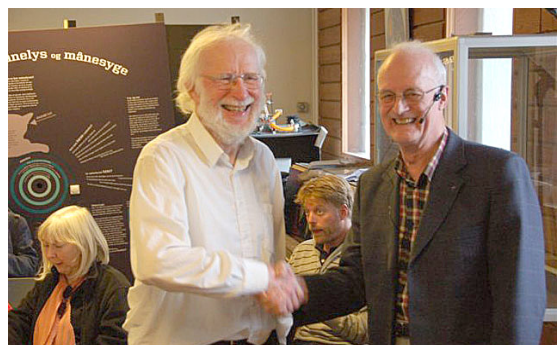

**Figur 10.** Peter Naur og Erik Høg mødtes i 2010.

### Litteratur

[1] Peter Naur citeres i Weekendavisen 3. sept. 2010, s. 12 i en artikel af Thore Bjørnvig og Karin Tybjerg om Brorfelde Observatorium og 50 års astrometri. Den 10. september fortsættes artiklen.

[2] C. Turon (2009), The Tycho-2 Catalogue, *Astronomy & Astrophysics* vol. **500**, p. 583,

[3] Information om Hipparcos (og link til Gaia), http://www.rssd.esa.int/Hipparcos

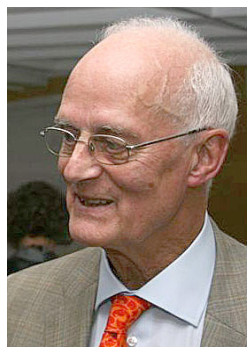

*Erik Høg*, dr. scient i astronomi. Har arbejdet ved Hamborg Observatoriet 1958-73 og ved Københavns Universitet 1953-58 og 1973-2002, hvor han gik på pension. Hans videnskabelige arbejde har været koncentreret indenfor måling af positioner, bevægelser og afstande af stjerner med højest mulig præcision. Han har bl.a. bidraget med design af to astrometriske satellitter: Hipparcos og Roemer/Gaia.

# Lectures on astrometry
*overview, handouts and abstracts*

*Erik Høg - Niels Bohr Institute, Copenhagen University - **email:** erik.hoeg@get2net.dk*

## No. 1:
### Astrometry Lost and Regained
### From a modest experiment in Copenhagen in 1925 to the Hipparcos and Gaia space missions

*Erik Høg*

ABSTRACT: Technological and scientific developments during the past century made a new branch of astronomy flourish, i.e. astrophysics, and resulted in our present deep understanding of the whole Universe. But this brought astrometry almost to extinction because it was considered to be dull and old-fashioned, especially by young astronomers. Astrometry is the much older branch of astronomy which performs accurate measurements of positions, motions and distances of stars and other celestial bodies. Astrometric data are of great scientific and practical importance for investigation of celestial phenomena and also for control of telescopes and satellites and for monitoring of Earth rotation. Our main subject is the development during the 20th century which finally made astrometry flourish as an integral part of astronomy through the success of the Hipparcos astrometric satellite, soon to be followed by the even more powerful Gaia mission. The Hipparcos mission approved in 1980 was based on photoelectric detectors measuring one star at a time. In 1992 CCD detectors were introduced in the Roemer mission proposal which could measure ten thousands of stars simultaneously, still in a rotating satellite performing a systematic scan of the entire sky. During 1993-97 an interferometric option, GAIA, was also studied, but the Roemer option with direct imaging on CCDs was much better and is therefore used in the Gaia mission.

++++++++++++++++++++++++++++++++++++++++++++++++++++++++++++++++++++++++

## No. 2:
### Hipparcos - Roemer - Gaia

*Erik Høg*

ABSTRACT: During the Hipparcos mission in September 1992, I presented a concept for using direct imaging on CCDs in scanning mode in a new and very powerful astrometric satellite, Roemer. The Roemer concept with larger aperture telescope for higher accuracy was developed by ESA and a mission was approved in 2000, expected to be a million times better than Hipparcos. The name Gaia mission reminds of an interferometric option also studied in the period 1993-97, and this period is a main subject of my presentation.

++++++++++++++++++++++++++++++++++++++++++++++++++++++++++++++++++++++

Handouts from presentations of lectures No. 1 and 2 in The Netherlands in January 2011:
**Astrometry Lost and Regained:**
**From a modest experiment in Copenhagen in 1925 to the Hipparcos and Gaia space missions**
Slides of 18 January 2011 shown in Amsterdam and Leiden:
www.astro.ku.dk/~erik/AstrometryLeiden.ppt.pdf

Contribution No.12         Lectures on astrometry

with links to documents on no. 19, 34, 38.  The lunch talk in Groningen was a short version hereof.
&

**Hipparcos-Roemer-Gaia:**
**From photoelectric astrometry with Hipparcos to CCD astrometry with Gaia**
Lecture on 21 January 2011 in ESTEC, Holland
www.astro.ku.dk/~erik/HippRoemerG.ppt.pdf   with links to documents on no. 16, 38, 46

Article: **Astrometry Lost and Regained:**   www.astro.ku.dk/~erik/AstromRega3.pdf
++++++++++++++++++++++++++++++++++++++++++++++++++++++++++++++++++++++++++

**Lecture No. 3**

## The Depth of Heavens - Belief and Knowledge during 2500 Years

   The lecture outlines the structure of the universe and the development of science during 5000 years, focusing on the distances in the universe and their dramatic change in the developing cultural environment from Babylon and ancient Greece to modern Europe.
   For Dante Alighieri (1265-1321) the spiritual cosmos contained the Heavens, Earth, and Hell, and it was compatible with the physical cosmos known from Aristotle (384-322 B.C.). Dante's many references in his Divine Comedy to physical and astronomical subjects show that he wanted to treat these issues absolutely correct. Tycho Brahe proves three hundred years later by his observations of the Stella Nova in 1572 and of comets that the spheres of heavens do not really exist. It has ever since become more and more difficult to reconcile the ancient ideas of a unified cosmos with the increasing knowledge about the physical universe.
   Ptolemy derived a radius of 20 000 Earth radii for the sphere of fixed stars. This radius of the visible cosmos at that time happens to be nearly equal to the true distance of the Sun, or 14 micro-light-years. Today the radius of the visible universe is a million billion (10 to the power 15) times larger than Ptolemy and Tycho Brahe believed.

**Lecture No. 4**:

## 400 Years of Astrometry: From Tycho Brahe to Hipparcos

Galileo Galilei's use of the newly invented telescope for astronomical observation resulted immediately in epochal discoveries about the physical nature of celestial bodies, but the advantage for astrometry came much later. The quadrant and sextant were pre-telescopic instruments for measurement of large angles between stars, improved by Tycho Brahe in the years 1570-1590. Fitted with telescopic sights after 1660, such instruments were quite successful, especially in the hands of John Flamsteed. The meridian circle was a new type of astrometric instrument, already invented and used by Ole Rømer in about 1705, but it took a hundred years before it could fully take over. The centuries-long evolution of techniques is reviewed, including the use of photoelectric astrometry and space technology in the first astrometry satellite, Hipparcos, launched by ESA in 1989. Hipparcos made accurate measurement of large angles a million times more efficiently than could be done in about 1950 from the ground, and it will soon be followed by Gaia which is expected to be another one million times more efficient for optical astrometry.
++++++++++++++++++++++++++++++++++++++++++++++++++++++++++++++++++++

# Short info about the four lectures

### Lecture No. 1.   45 minutes
### Astrometry Lost and Regained
**From a modest experiment in Copenhagen in 1925
to the Hipparcos and Gaia space missions**

   The lecture has been developed over many years and was held in, e.g., Copenhagen, Vienna, Bonn, Düsseldorf, Vilnius, Oslo, Nikolajev, Poltava, Kiev, Thessaloniki, Ioannina, Athens, Rome, Madrid, Washington, and Charlottesville - since 2007 in PowerPoint.  Revised in 2009 and with the new title *Astrometry Lost and Regained* it was held in Heidelberg, Sct. Petersburg, Rio de Janeiro, Morelia, Mexico City, Beijing, Montpellier, Groningen, Amsterdam, and Leiden.

---

### Lecture No. 2.   45 minutes
### Hipparcos - Roemer - Gaia

   **The lectures briefly outlines the development of photoelectric astrometry culminating with the Hipparcos mission. Development of the Gaia mission beginning in 1992 is followed in detail.**
   The lecture has been held since 2010 in Toulouse and at ESTEC in Holland.
___________________________________________________________________________

### Lecture No. 3:   45 minutes. Suited for a broad audience, including non-astronomers
### The Depth of Heavens - Belief and Knowledge during 2500 Years

   **The lecture outlines the structure of the universe and the development of science during 5000 years, focusing on the distances in the universe and their dramatic change in the developing cultural environment from Babylon and ancient Greece to modern Europe.**
   The lecture was first held in 2002, and since 2007 in PowerPoint. Held in Copenhagen, Vilnius, Nikolajev, Athens, Catania, Madrid, and Paris.
       Handouts at:  www.astro.ku.dk/~erik/DepthHeavens2.pdf
    and   www.astro.ku.dk/~erik/DepthHeavens.pdf

   **An article with the same title as the lecture** appeared in Europhysics News (2004) Vol. 35 No.3.
Here slightly updated, 2004.02.20:   www.astro.ku.dk/~erik/Univ7.5.pdf

---

### Lecture No. 4.   45 or 30 minutes.
### 400 Years of Astrometry: From Tycho Brahe to Hipparcos

   **The four centuries of techniques and results are reviewed, from the pre-telescopic era until the use of photoelectric astrometry and space technology in the first astrometry satellite, Hipparcos, launched by ESA in 1989.**
   The lecture was presented as invited contribution to the symposium at ESTEC in September 2008: **400 Years of Astronomical Telescopes: A Review of History, Science and Technology.** The report to the proceedings is included as No. 8 among the "Contributions to the history of astrometry ".
   It was later held in Sct. Petersburg, Rio de Janeiro, and Morelia.
   +++++++++++++++++++++++++++++++++++++++++++++++++++++++++++++